\documentclass[a4paper,twocolumn,10pt,accepted=2024-07-04]{quantumarticle}
\pdfoutput=1
\usepackage{bm,bbm,xcolor}
\usepackage{amsmath, amssymb}
\usepackage{mathrsfs} 
\usepackage{dsfont}
\usepackage{braket,leftindex,enumerate}
\usepackage[numbers,sort&compress]{natbib}
\usepackage[colorlinks=true,breaklinks=true,allcolors=blue]{hyperref}

\definecolor{giovanna}{rgb}{.75,0,1}

\definecolor{emma}{rgb}{.0,.0,1}

\definecolor{hannes}{rgb}{.89,.0,.13}

\definecolor{raphael}{rgb}{.5,.6,.13}

\definecolor{luigi}{rgb}{.2,.6,.2}

\definecolor{changes}{rgb}{1.0, 0.0, 0.25}

\begin{document}
\title{Adiabatic quantum trajectories in engineered reservoirs}

\author{Emma C. King}
\affiliation{Theoretische  Physik,  Universit\"at  des  Saarlandes,  D-66123  Saarbr\"ucken,  Germany}
\orcid{0000-0002-6696-3235}

\author{Luigi Giannelli}
\affiliation{Dipartimento di Fisica e Astronomia ``Ettore Majorana'', Università di Catania, Via S. Sofia 64, 95123 Catania, Italy}
\affiliation{CNR-IMM, UoS Università, 95123 Catania, Italy}
\affiliation{INFN Sezione di Catania, 95123 Catania, Italy}
\orcid{0000-0001-9704-7304}

\author{Rapha\"el Menu}
\affiliation{Theoretische  Physik,  Universit\"at  des  Saarlandes,  D-66123  Saarbr\"ucken,  Germany}
\orcid{0000-0001-7641-9922}

\author{Johannes N. Kriel}
\affiliation{Institute of Theoretical Physics, Stellenbosch University, Stellenbosch 7600, South Africa}
 \orcid{0000-0002-3852-7823}

\author{Giovanna Morigi}
\affiliation{Theoretische  Physik,  Universit\"at  des  Saarlandes,  D-66123  Saarbr\"ucken,  Germany}
\orcid{0000-0002-1946-3684}

\begin{abstract}
We analyze the efficiency of protocols for adiabatic quantum state transfer assisted by an engineered reservoir. The target dynamics is a quantum trajectory in the Hilbert space and is a fixed point of a time-dependent master equation in the limit of adiabatic dynamics. We specialize to quantum state transfer in a qubit and determine the optimal schedule for a class of time-dependent Lindblad equations. The speed limit on state transfer is extracted from a physical model of a qubit coupled to a reservoir, from which the Lindblad equation is derived in the Born-Markov limit. Our analysis shows that the resulting efficiency is comparable to the efficiency of the optimal unitary dynamics. Numerical studies indicate that reservoir-engineered protocols could outperform unitary protocols outside the regime of the Born-Markov master equation, namely, when correlations between the qubit and reservoir become relevant. Our study contributes to the theory of shortcuts to adiabaticity for open quantum systems and to the toolbox of protocols of the NISQ era. 
\end{abstract}
\maketitle 

\section{Introduction}
Widely employed strategies in quantum technologies make use of adiabatically steering the trajectory of a quantum system in the Hilbert space \cite{Albash_Lidar18,Vitanov:2017,Santoro_Tosatti06}. In fact, adiabatic processes posses an inherent robustness to parameter variations, provided the control parameters of the system are varied sufficiently slowly to satisfy the adiabaticity constraint. Adhering to this constraint, however, comes at the expense of long operation times, during which the system is susceptible to accumulated effects of decoherence, excitation transitions and particle loss. For this reason, the characteristic timescale associated with these stochastic processes, typically referred to as noise, sets a bound on the operation time of the adiabatic protocol, and thus on the maximal fidelity that can be achieved~\cite{Keck_etal17,Preskill2018quantumcomputingin}. 

This stumbling block might be overcome with shortcuts to adiabaticity (STA)~\cite{Guery-Odelin_etal19,delCampo:2019}. STA is a class of methods and concepts that aim to find fast routes to the final results of slow, adiabatic protocols, thereby also reducing the detrimental impact of decoherence. This promotes a positive step towards the realization of fast and accurate quantum processes. STA relates to and overlaps partially with optimal control theory \cite{Guery-Odelin_etal19,Hoff_etal_2012,Koch_etal_22,Giannelli_etal_2022}. The minimal time needed for implementing the protocol is bound by the quantum speed limit intrinsic to the system's dynamics \cite{Caneva:2009,Deffner_2017}. Following a somewhat different approach, strategies based on quantum reservoir engineering \cite{Poyatos_etal_1996} make use of non-unitary, completely positive and trace preserving (CPTP)  maps, which are tailored such that the target state is a fixed point \cite{Kraus_etal_2008,Roy-Gefen:2020}. These maps are generated by master equations, with the non-unitary features giving rise to an arrow of time. The characteristic timescale of the corresponding protocol is bound by a quantum speed limit determined by the interplay between coherent and incoherent dynamics \cite{delCampo:2013,Funo_2019,Fogarty:2020}. 

Quantum reservoir engineering has been discussed for a variety of applications, from quantum state preparation \cite{Poyatos_etal_1996,Pielawa:2007,Krauter_etal_2011,Lin_NIST_2013,Morigi:2015} to quantum computing \cite{Verstraete:2009,Barreiro:2011} and quantum communication \cite{Vollbrecht:2011,Blycka:2014}. Recent works developed a general framework for achieving shortcuts to adiabaticity assisted by quantum reservoir engineering, which we here dub by the acronym STARE. These protocols consist in using non-unitary, time-dependent Lindblad dynamics in order to suppress diabatic transitions \cite{Vacanti_2014,Alipour:2020shortcutsto}. In this spirit, Ref.~\cite{Menu_etal22} proposed a protocol that achieves a shortcut to adiabaticity in a Landau-Zener model by means of a quantum auxiliary system (meter), performing projective measurements into the target adiabatic trajectory. Using this scheme, the probability of a diabatic transition in a given time can be reduced by up to one order of magnitude with respect to the Landau-Zener transition. This result makes this protocol interesting for applications to quantum technologies, for quantum state preparation in quantum simulators \cite{Venuti:2017} and prominently for quantum computing, where adiabatic transfer in a qubit is quintessential for the adiabatic quantum search \cite{Wild_etal16}. In order to assess its potential, it is necessary to identify the ultimate speed limit when compared to the speed limit of the protocol based on the coherent dynamics \cite{Roland_Cerf02}.

In this work we assess the speed limit of STARE protocols for a qubit based on the time-dependent Born-Markov Lindblad master equation of Refs.~\cite{Avron_etal10,Menu_etal22}. Our analysis allows us to identify the optimal schedule and, subsequently, the lower bound on the transfer time that this type of STARE protocol can achieve. 

The paper is organized as follows. In Sec.~\ref{Sec:STARE} we summarize the basic idea of the STARE protocol considered here. In Sec.~\ref{s:two_level_system} we establish the foundation of STARE for quantum state transfer in a qubit: We first review the unitary adiabatic dynamics, and then identify the requirements for its extension to the setting of open quantum systems. In Sec.~\ref{s:adiabatic_Lindblad_dynamics} we consider a specific class of Lindblad master equations fulfilling the requirements of Sec.~\ref{s:two_level_system}, and derive the optimal schedule for the adiabatic transfer. The comparison between the maximum efficiency of the unitary and STARE protocols is performed in Sec.~\ref{Sec:efficiency}. In Sec.~\ref{s:speed_limit} we assess the ultimate speed limit of the STARE protocol based on the Born-Markov Lindblad master equation by means of a microscopic model, from which the applicable master equation can be derived. We use this model to numerically explore the efficiency in the regime beyond the Born-Markov limit for specific implementations of experimental relevance. The conclusions are drawn in Sec.~\ref{s:summary}. The appendices provide details of the derivations in Sections \ref{s:adiabatic_Lindblad_dynamics} and \ref{s:speed_limit}.

\section{Quantum reservoir engineering for adiabatic transfer} \label{Sec:STARE}
Approaches based on quantum reservoir engineering design CPTP maps $\mathcal M$ for which the target density operator $\varrho_T$ is a fixed point: $\mathcal M\varrho_T=\varrho_T$. This is often implemented by means of a master equation of Lindblad form
for the density operator $\varrho$, 
\begin{equation}
\partial_t\varrho=\mathcal L\varrho\,,     
\end{equation}
with $\mathcal L$ a Liouville superoperator (Lindbladian) which generates the desired map, such that $\varrho(t)={\rm e}^{t\mathcal{L}}\varrho(0)$ \cite{Kraus_etal_2008,Morigi:2015,Albert_etal16}. By ensuring that $\varrho_T$ is the unique stationary state, namely, $\mathcal L\varrho_T=0$, the system will dynamically be steered towards the target state. The advantage of protocols based on non-unitary maps generally derive from their relative robustness against parameter fluctuations. Moreover, since the target state is the stationary state, the required fidelity is reached on a timescale set by the relaxation of the system, thus negating the need for precise control of the evolution time. Finally, the impact of undesired, incoherent dynamics can, at least partially, be compensated for by the designed Lindbladian.

The STARE protocol we consider extends this approach to designing quantum trajectories by identifying time-dependent non-unitary dynamics $\mathcal L(t)$ such that the solution of the master equation tends to the target trajectory $\varrho_T(t)$ in the adiabatic limit. To this end, we generalize the condition of quantum coherent adiabatic dynamics and require that the quantum system is adiabatically transported along the trajectory $\varrho_T(t)$ satisfying the equation \cite{Albert_etal16}
\begin{equation}
\label{Eq:L(t)}
{\mathcal L(t)}\varrho_T(t)=0
\end{equation}
at each instant of time. The target trajectory is therefore the instantaneous right eigenvector of ${\mathcal L(t)}$ with eigenvalue zero. In this work we verify the conditions on the schedule for which $\varrho_T(t)$ is a solution of the master equation. We identify the optimal schedule of a class of Lindbladian for a paradigmatic problem, namely, adiabatic transfer in a two-level system (a qubit). This permits us to benchmark the efficiency of the STARE protocol with respect to protocols based on unitary adiabatic transport. 

\section{Adiabatic transfer in a qubit} \label{s:two_level_system}
A two-level quantum system is arguably the simplest non-trivial object in quantum theory, and is of great practical importance. 
Indeed, the Hamiltonian dynamics of adiabatic quantum state transfer in a two-level system is a paradigmatic model for the description of state preparation of a qubit, and adiabatic quantum searches in a database \cite{Wild_etal16,Albash_Lidar18}. In this section, we review the basics of adiabatic transfer between two orthogonal states of a qubit, and highlight an optimized schedule for this process. We then lay the foundation for the identification of the corresponding STARE protocol. While the results of this section are widely available in the literature, the notation and context established here underpin our main results on the optimization of the non-unitary dynamics in Sec.~\ref{s:adiabatic_Lindblad_dynamics}.

\subsection{Unitary dynamics}\label{ss:LZ_and_RC_sweeps}
\begin{figure}[t]
    \centering
    \includegraphics[width=0.48\columnwidth]{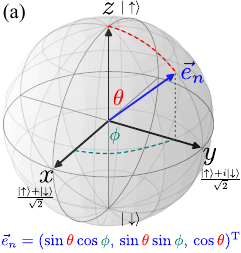}
    \includegraphics[width=0.48\columnwidth]{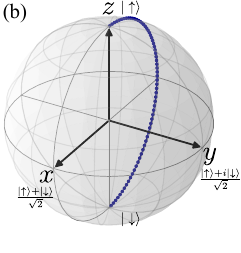}
    \caption{(a) Illustration of the Bloch sphere in $\mathbb{R}^3$, with the eigenstates of the Pauli matrix $\hat{\sigma}_z$ indicated by $\ket{\uparrow}$ and $\ket{\downarrow}$ at the $z$-axis poles. The unit vector $\vec{e}_n$ represents a pure state and is parameterized by the polar angle $\theta$ and the azimuthal angle $\phi$. (b) The optimal quantum trajectory $\ket{\psi(t)}_T$ of Eq.~\eqref{eq:RC_protocol_LZ_model} connecting the states $\vert\psi(t_i)\rangle_T=\ket{\uparrow}$ and the target state $\vert\psi(t_f)\rangle_T=\ket{\downarrow}$ with the minimal infidelity for a fixed transfer time.} 
    \label{fig:bloch_sphere}
\end{figure}

\noindent {\it Definitions.} We begin with a brief overview of the unitary dynamics of a qubit, generated by a time-dependent Hamiltonian $H(t)$. Here the evolution of $\varrho(t)$ is governed by the von Neumann equation $\partial_t\varrho(t)={\mathcal L}(t)\varrho(t)=[H(t),\varrho(t)]/i$ (with the convention $\hbar=1$). This dynamics preserves the purity of $\varrho(t)$, and, for pure initial states, the state of the qubit is therefore described by a vector $\ket{\psi(t)}$ in the system's two-dimensional Hilbert space $\mathcal{H}$. Choosing as a basis for $\mathcal{H}$ the eigenstates $\ket{\uparrow}$ and $\ket{\downarrow}$ of the Pauli matrix $\hat{\sigma}_z$, for which $\hat{\sigma}_z\ket{\uparrow}=+\ket{\uparrow}$ and $\hat{\sigma}_z\ket{\downarrow}=-\ket{\downarrow}$, we write the qubit's state as 
\begin{equation}
    \ket{\psi(t)} =  c_\uparrow(t) \ket{\uparrow} + c_\downarrow(t) \ket{\downarrow},
\end{equation}
where $c_{j}(t)=\langle j|\psi(t)\rangle$ with $j=\,\uparrow,\downarrow$ are expansion coefficients satisfying $|c_\uparrow(t)|^2+|c_\downarrow(t)|^2=1$. Turning now to the Hamiltonian, we first consider the general form 
\begin{equation}\label{eq:generic_two_level_system}
    H(t) = a_0(t) \mathbb{I} + \vec{a}(t) \cdot \vec{\sigma}\,,
\end{equation}
where $\vec{a}(t)=(a_x(t),\, a_y(t),\, a_z(t))^\text{T}\in\mathbb{R}^3$ and $\vec{\sigma}=(\hat{\sigma}_x,\, \hat{\sigma}_y,\, \hat{\sigma}_z)^\text{T}$ is the Pauli-matrix vector operator. The unit direction vector $\vec{e}_n(t)=\vec{a}(t)/|\vec{a}(t)|$ may be identified with a point on the Bloch sphere, and parameterized by the angles $\theta\in[0,\pi)$ and $\phi\in[0,2\pi)$ as $\vec{e}_n=(\sin\theta \cos\phi,\, \sin\theta\sin\phi,\, \cos\theta)^\text{T}$. See Fig.~\ref{fig:bloch_sphere}(a) for an illustration of these notations. The Hamiltonian~\eqref{eq:generic_two_level_system} is then diagonal in the eigenbasis 
\begin{subequations} \label{eq:generic_eigenstates_2Level}
\begin{align}
    \ket{+}_t &= +\cos[\theta(t)/2] \ket{\uparrow} + e^{i\phi} \sin[\theta(t)/2]\ket{\downarrow},\\
    \ket{-}_t &= -\sin[\theta(t)/2] \ket{\uparrow} + e^{i\phi} \cos[\theta(t)/2]\ket{\downarrow},
\end{align}
\end{subequations}
with \mbox{$\tan[\theta(t)]=\sqrt{a_x^2(t)+a_y^2(t)}/a_z(t)$} and $\tan[\phi(t)]=a_y(t)/a_x(t)$. The associated eigenenergies are
\begin{equation} \label{eq:generic_energies_2Level}
   E_\pm(t) = a_0(t) \pm |\vec{a}(t)|\,.
\end{equation}
We denote the instantaneous gap in the spectrum of $H(t)$ by $\Delta(t)\equiv E_+(t)-E_-(t)$. With our notation, $\Delta(t)=2|\vec{a}(t)|$. As will be seen, the minimum value $\Delta_{\min}$ of this gap is instrumental in bounding the accuracy and speed of adiabatic transfer in this system.\\

\noindent {\it The quantum trajectory.} Now consider the unitary adiabatic transfer of the qubit's state along a smoothly varying target trajectory $\ket{\psi(t)}_T$ which connects the states $\ket{\psi(t_i)}_T=\ket{\uparrow}$ and $\ket{\psi(t_f)}_T=\ket{\downarrow}$ over a time interval $[t_i,t_f]$. To realize this transfer, we must identify a quantum evolution
\begin{equation}\label{eq:unitary_evolution}
i\partial_t\ket{\psi(t)}=H(t)\ket{\psi(t)},
\end{equation}
for which $\ket{\psi(t)}_T$ is an instantaneous eigenstate of $H(t)$, satisfying $H(t)\ket{\psi(t)}_T=E(t)\ket{\psi(t)}_T$ for all \mbox{$t\in[t_i,t_f]$}. Taking the initial qubit state as \mbox{$\ket{\psi(t_i)}=\ket{\uparrow}$}, the evolution of $\ket{\psi(t)}$ will then tend to the target trajectory $\ket{\psi(t)}_T$ in the limit of vanishing rate of variation of $H(t)$ \cite{Born_Fock1928,Kato50}. 

\begin{figure*}[ht]
    \centering
    \includegraphics[width=0.68\columnwidth]{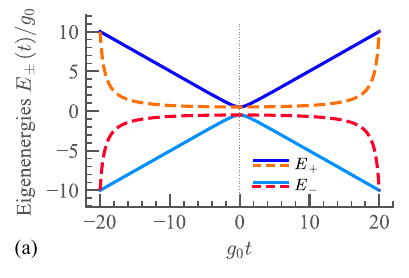}
    \includegraphics[width=0.68\columnwidth]{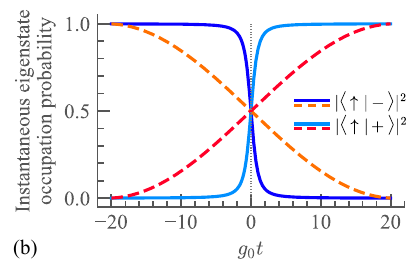}
    \includegraphics[width=0.68\columnwidth]{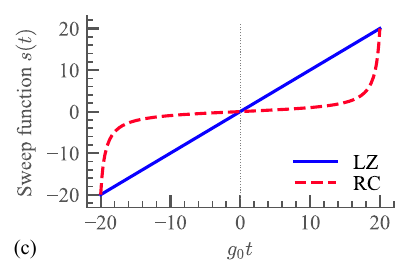}
    \caption{(a) The instantaneous eigenenergies $E_\pm(t)$ and (b) the occupation probabilities $|\langle \uparrow|\pm\rangle|^2$  of the instantaneous eigenstates $\ket{\pm}_t$ of the two-level Hamiltonian in Eq.~\eqref{eq:LZ_Hamiltonian} as a function of the rescaled time $g_0t$ and for two different sweep functions: the linear sweep function (LZ) with $s(t)=\epsilon t$ (solid line) and the optimal sweep function of Eq.\ \eqref{eq:RC_protocol_LZ_model} (dashed line). The dotted vertical line indicates the time $t=0$ where the energy gap is minimum. Subplot (c) displays the functional dependence of the two sweep functions on time.} 
    \label{fig:EVs_schedule}
\end{figure*}

Now, we turn our attention to identifying a suitable Hamiltonian in Eq.~\eqref{eq:unitary_evolution}. We choose $a_0=0$ and \mbox{$\vec{a}(t) = (g_0/2,\, 0,\, s(t)/2)^\text{T}$} in Eq.~\eqref{eq:generic_two_level_system}, which yields
\begin{equation} \label{eq:LZ_Hamiltonian}
    H(t) = \frac{s(t)}{2}\hat{\sigma}_z +\frac{g_0}{2} \hat{\sigma}_x\,,
\end{equation}
with $s(t)$ a monotonically increasing function of time, and $g_0>0$. By taking $s(t_i)<0<s(t_f)$ and $|s(t_i)|,|s(t_f)|\gg g_0$, the instantaneous ground state $\ket{-}_t$ in Eq.~\eqref{eq:generic_eigenstates_2Level} will satisfy $\ket{-}_{t_i}\approx\ket{\uparrow}$ and $\ket{-}_{t_f}\approx\ket{\downarrow}$, and thereby provide a realization of the desired target trajectory. We refer to $s(t)$ as the \textit{sweep function}, since it controls the sweeping rate of the Hamiltonian. Here $g_0$ couples the $\ket{\downarrow}$ and $\ket{\uparrow}$ states, and determines the minimal gap $\Delta_{\rm min}=g_0$ by lifting the degeneracy at the avoided level-crossing at $t=0$. See Fig.~\ref{fig:EVs_schedule}(a) and \ref{fig:EVs_schedule}(b) for the instantaneous eigenenergies and occupation probabilities, respectively, and for two different sweep functions $s(t)$.\\

\noindent {\it Adiabaticity.} The rate of variation of $s(t)$, and how this rate compares to the minimal gap $\Delta_{\rm min}=g_0$, will determine the adiabaticity of the dynamics generated by $H(t)$ in Eq.~\eqref{eq:LZ_Hamiltonian}, and therefore the extent to which $\ket{\psi(t)}$ 
"follows" the target trajectory $\ket{\psi(t)}_T$. This directly impacts the efficiency of the transfer protocol. We will assess the latter by means of the infidelity $\mathcal{I}$, which is the probability that the upper energy branch is populated at the end of the process, 
\begin{equation}\label{eq:tunnelingProb_LZ_general}
    \mathcal{I}(t_i,\,t_f) = |\leftindex_{t_f}{}\langle +|\, \hat{U}(t_i,t_f) \ket{-}_{t_i}|^2\,,
\end{equation}
where $\hat{U}(t_i,t_f)$ is the unitary operator evolving the system according to Eq.~\eqref{eq:unitary_evolution}. 

For a linear sweep at constant rate $\epsilon$, $s(t)=\epsilon t$, the dynamics is that of the paradigmatic Landau-Zener problem. In the limits \mbox{$t_i\rightarrow -\infty$} and $t_f\rightarrow +\infty$, the infidelity is then given by the analytic expression~\cite{Zener32,Landau65,Landau32}
\begin{equation}\label{eq:LZ_formula}
    \mathcal{I}_\mathrm{LZ} = \lim_{\substack{t_i\rightarrow -\infty\\t_f\rightarrow +\infty}} \mathcal{I}(t_i,t_f) = e^{-\pi g_0^2/(2\epsilon)}\,,
\end{equation}
see \cite{Mullen:1989} for a detailed analysis of the tunneling timescale. The competition between the square of the minimal energy gap $g_0$ and the sweep rate $\epsilon$ is apparent in the expression above. Adiabatic dynamics occurs when $g_0\gg \sqrt{\epsilon}$, when the minimum energy gap $g_0$ is then sufficiently large to suppress transitions out of the instantaneous ground state. We remark that for finite transfer times $(T=t_f-t_i)$ and symmetric sweeps $(t_f=-t_i$), the asymptotic behavior of the infidelity scales as $\mathcal{I}_\mathrm{LZ}\sim g_0^2/(\epsilon^4 T^6)$ for sufficiently small $\epsilon$ \cite{Grandi_Polkovnikov10}.

Roland and Cerf \cite{Roland_Cerf02} derived an optimal protocol which minimizes the infidelity for a given transfer time or, equivalently, decreases the transfer time for a given infidelity. The sweep function is
\begin{equation}\label{eq:RC_protocol_LZ_model}
    s(t) = g_0 \tan\left[\frac{2t}{T} \arctan\left(\frac{\epsilon T}{2g_0}\right) \right],
\end{equation}
which satisfies $s(t_f)= \epsilon T/2$ and $s(t_i)= -\epsilon T/2$. The target trajectory on the Bloch sphere is displayed in Fig.~\ref{fig:bloch_sphere}(b), and the functional behavior of the schedule  with time is shown in Fig.~\ref{fig:EVs_schedule}(c). One can observe that $s(t)$ now varies slowly in the vicinity of the avoided level-crossing at $s(0)=0$. Meanwhile, away from this crossing, the gradient of $s$ rapidly increases, facilitating faster adiabatic transfer. The Roland and Cerf protocol, together with the linear Landau-Zener protocol, have been experimentally studied in Refs.~\cite{BasonNP2012highfidelity,MalossiPRA2013quantum}. These works also discuss extensions to transitionless driving protocols and their robustness against parameter variations.\\

\noindent {\it Dimensionless parameters.} In order to quantify the regime of adiabatic dynamics, we introduce the adiabaticity parameter 
\begin{equation} \label{eq:dimensionless_parameters_a}
    a=g_0T\,.
\end{equation}
If the start and end points $\ket{\psi(t_i)}_T$ and $\ket{\psi(t_f)}_T$ of the target trajectory, and therefore the values of $s(t_i)$ and $s(t_f)$ in Eq.~\eqref{eq:LZ_Hamiltonian}, are held fixed, the adiabatic regime requires $a\gg1$. In fact, an expansion of $\ket{\psi(t)}$ in orders of $1/a$ will yield, at leading order, the target trajectory $\ket{\psi(t)}_T$, with non-adiabatic corrections at higher orders \cite{Kato50,Grandi_Polkovnikov10}. The adiabaticity parameter also appears naturally in the Schr\"odinger equation upon rescaling energies by $g_0$ and time by $T$. In terms of the dimensionless time 
\begin{equation} \label{eq:dimensionless_time}
\tau=(t-t_i)/T\,;\,\,\,\tau\in[0,1],
\end{equation}
the dynamics of the qubit state $\ket{\psi(\tau)}$ is then governed by the Hamiltonian 
\begin{equation} \label{eq:HqGeneral}
    H_{q}(\tau) = \frac{a}{2}\left[d(\tau)\hat{\sigma}_z +\hat{\sigma}_x \right]\,,
\end{equation}
where $d(\tau)=s(t)/g_0$. We define the shorthand \mbox{$d_i=s(t_i)/g_0$} and $d_f=s(t_f)/g_0$. It will be convenient to write $d(\tau)$ in the form
\begin{equation} \label{eq:sweep_function_dimensionless_parameters}
    d(\tau) = q(\tau) d_f + [1-q(\tau)]d_i\,, 
\end{equation}
where the schedule $q(\tau)$ is monotonically increasing, and satisfies $q(0)=0$ and $q(1)=1$. The sweep function of Eq.~\eqref{eq:RC_protocol_LZ_model} is then equivalent to the choice of schedule 
\begin{equation} \label{eq:RC_schedule_dimensionless}
    q_\mathrm{RC}(\tau) = \frac{ \tan\left[\tau \arctan d_f +(1- \tau) \arctan d_i \right] -d_i}{d_f-d_i} \,.
\end{equation}
It turns out that for the linear Landau-Zener sweep a different choice of the dimensionless adiabaticity parameter is more appropriate. This choice also enables one to identify the scaling of the transfer time with the minimum gap $g_0$. We report these considerations in Appendix~\ref{app:LZ}. Before concluding, we emphasize that with this choice of the dimensionless variables, and by keeping $d_i$ and $d_f$ constant, the transfer time $T$ now determines the speed of the transformation: Increasing $T$ implies decreasing the speed of the sweep.

\subsection{Incoherent dynamics}\label{Sec:IncoherentDynamics}
We now turn to the task of adapting the formalism of the previous section to the setting of non-unitary dynamics, where the target trajectory corresponds to a density operator $\varrho_T(t)$. In accordance with the discussion in Section~\ref{Sec:STARE}, we need to design a Liouville superoperator fulfilling the condition $\mathcal L(t)\varrho_T(t)=0$. \\

\noindent {\it The Liouville superoperator.} Let us consider a Lindblad superoperator of the general form \cite{Breuer_Petruccione07}
\begin{align}
\label{eq:Liouvillian}
&{\mathcal L}(t)\varrho(t)=-i[H(t),\varrho(t)]\\
&+\sum_{\alpha}\gamma_\alpha(t)\left([L_\alpha(t),\varrho(t) L_\alpha^\dagger(t)]+[L_\alpha(t)\varrho(t), L_\alpha^\dagger(t)]\right),\nonumber
\end{align}
with $L_{\alpha}(t)$ a time-dependent jump operator over a finite-dimensional Hilbert space $\mathcal{H}$ and where $\gamma_\alpha(t)$ is real and positive. In order to fulfil Eq.~\eqref{Eq:L(t)}, we require that the Hamiltonian and the jump operators commute with the target trajectory $\varrho_T(t)$, such that 
\begin{equation}
\label{eq:commutator}
[H(t),\varrho_T(t)]=0 \quad {\rm and} \quad [L_\alpha(t),\varrho_T(t)]=0 
\end{equation}
hold for all $t\in[t_i,t_f]$. These restrictions, together with requiring the hermiticity of the jump operators $L_\alpha$, are sufficient to ensure that the condition $\mathcal L(t)\varrho_T(t)=0$ is satisfied. In what follows we denote the Lindbladian of Eq.~\eqref{eq:Liouvillian} whose jump operators fulfil the condition of Eq.~\eqref{eq:commutator} by STARE Lindbladian. Its action is an effective dephasing that suppresses transitions out of the target trajectory, see Fig.~\ref{fig:bloch_sphere}(b). For a qubit, the STARE Lindbladian takes the form
\begin{align}\label{eq:dephasingME_two_level:0} 
     &\mathcal{L}(t)\varrho(t) = -i[H(t),\varrho(t)] \\
     &+\gamma(t) \left(\left[\Pi_z (t), \varrho(t) \Pi_z (t)\right] + \left[\Pi_z(t)\varrho(t),\Pi_z(t)\right]\right),\nonumber
\end{align}
with $\Pi_z(t)=(P_+ (t)-P_- (t))/2$ and where \mbox{$P_\pm(t) = |\pm\rangle_{t}\langle\pm|$} \eqref{eq:generic_eigenstates_2Level} are the projectors onto the instantaneous excited and ground state of $H(t)$ \eqref{eq:LZ_Hamiltonian}. Expanding the commutators in Eq.~\eqref{eq:dephasingME_two_level:0} and simplifying, $\mathcal{L}(t)$ reduces to the form
\begin{align}\label{eq:dephasingME_two_level} 
     \mathcal{L}(t)\varrho(t) &= -i[H(t),\varrho(t)] -\gamma(t)\mathcal{D}(t)\varrho(t)\,,\\
     \mathcal{D}(t)\varrho(t) &= \left[P_+ (t) \varrho(t) P_- (t) + P_- (t)\varrho(t) P_+ (t)\right].\nonumber
\end{align}
The incoherent dynamics suppresses transitions between the instantaneous eigenstates of the Hamiltonian, therefore stabilizing the adiabatic trajectory.\\
 
\noindent {\it Dimensionless parameters.} For later analysis it is useful to rewrite the master equation in a dimensionless form. The incoherent dynamics introduces the dephasing rate $\gamma(t)$, which gives rise to an additional dimensionless parameter
\begin{equation} \label{eq:dimensionless_parameters_b}
b(\tau)=\gamma(\tau) T\,,
\end{equation}
which serves as a measure of the strength of the incoherent dynamics. The master equation based on the STARE Lindbladian~\eqref{eq:dephasingME_two_level} may be recast as
\begin{equation} \label{eq:Lindbladian_dimensionless}
        \partial_\tau\varrho(\tau) = -i \left[H_q(\tau),\varrho(\tau)\right] - b(\tau)\mathcal{D}[q(\tau)]\varrho(\tau)\,,
\end{equation}
with $H_q(\tau)$ as given in Eq.~\eqref{eq:HqGeneral}. 
Note that the explicit dependence on the dimensionless time $\tau=(t-t_i)/T$ enters on the right-hand side of the master equation~\eqref{eq:Lindbladian_dimensionless} only through the schedule $q(\tau)$.

Adiabatic transport by means of Lindbladian \eqref{eq:dephasingME_two_level} has been extensively studied in Refs.~\cite{Avron_etal10,Avron_etal11}. In what follows, we identify a schedule that optimizes this dynamics. Thereby, we determine the speed limit of adiabatic transfer. To this end, in the next section we review the theory of adiabatic transport by Avron and collaborators \cite{Avron_etal10,Avron_etal11,Avron_etal12}. We then apply it to derive the optimal schedule in the limit where the incoherent contribution to the dynamics is dominant.

\section{Adiabatic Lindblad dynamics} \label{s:adiabatic_Lindblad_dynamics}
In this section, we present results of an analytic analysis of the adiabatic transport generated by the STARE Lindbladian of Eq.~\eqref{eq:dephasingME_two_level}. Specifically, we perform an adiabatic expansion of the density matrix $\varrho(t)$ evolving according to Eq.~\eqref{eq:Lindbladian_dimensionless}. The leading-order term in this expansion is precisely the desired target trajectory $\varrho_T(t)$, while the higher-order terms introduce non-adiabatic corrections. From this result we then derive an analogous expansion for the infidelity, which we use to determine the optimal schedule for adiabatic transfer with the STARE Lindbladian. 

\subsection{The parallel transport propagator}\label{app:ss:parallel transport}
Our expansion of the density matrix is based on the adiabatic theory for Lindblad generators developed in Refs.~\cite{Avron_etal87,Avron_etal12_CMP}, see also Ref.\ \cite{Venuti:2016}. As set out in Section~\ref{Sec:IncoherentDynamics}, we have in mind the target trajectory $\varrho_T(\tau)=P_-(\tau)$, which connects the ground states of the Hamiltonian in Eq.~\eqref{eq:HqGeneral} between $\tau=0$ ($t=t_i$) and $\tau=1$ ($t=t_f$), with the initial condition $\varrho(0)=P_-(0)$. 
For this purpose, we introduce the instantaneous right eigenvectors of the STARE Lindbladian $\mathcal{L}$ as $S_{ij}=|i\rangle\langle j|$, $i,j\in\{+,-\}$. The operators $S_{ij}$ include the projectors $P_{j}=S_{jj}$, and satisfy the eigenvalue equation $\mathcal{L}S_{ij}=\lambda_{ij}S_{ij}$, where $\lambda_{ij}$ is the corresponding instantaneous eigenvalue with $\lambda_{ij}=0$ for $i=j$. Note that these right eigenvectors form a basis for the Banach space $\mathcal{B}$ of operators on the system's Hilbert space $\mathcal{H}$. With this, the kernel (nullspace) of $\mathcal{L}$, defined as $ \text{ker}(\mathcal{L}) = \{S\in\mathcal{B}\,|\,\mathcal{L}S=0\}\subset\mathcal{B}$, is spanned by $S_{ii}=P_i$, with $i=\pm$ \cite{Avron_etal12_CMP}. Meanwhile, the range (image) of $\mathcal{L}$, namely $\text{ran}(\mathcal{L}) = \{\mathcal{L}S\,|\,S\in\mathcal{B}\}\subset \mathcal{B}$,
is spanned by $S_{ij}$ for $i\neq j$ \cite{Avron_etal12_CMP}. In our case, $\mathcal{B} = \text{ker}(\mathcal{L})\oplus \text{ran}(\mathcal{L})$ and the relevant superoperator projections on $\mathcal{B}$ are
\begin{equation} \label{eq:PandQ_superprojectors}
    \mathcal{P}\varrho = \sum_{j=\pm} P_j \varrho P_j\,,\quad \mathcal{Q}\varrho =\sum_{\mathclap{\substack{j\neq k\\j,k=\pm}}} P_j \varrho P_k\,, \quad \mathcal{P}+\mathcal{Q} = \mathbb{I}\,.
\end{equation}
Here $\mathcal{P}$ projects onto the time-dependent manifold of stationary states, $\text{ker}(\mathcal{L})$. A system evolving adiabatically will remain in the manifold, i.e.\ there are no transitions from the family of projections $P\in\mathcal{P}$ to the bundle of complementary projections $\mathcal{Q}$ \cite{Avron_etal12_CMP}, and $P$ is said to undergo \textit{parallel transport} \cite{Fraas_Hanggli17,Avron_etal12_CMP}. This dynamics is the solution to the evolution equation \cite{Albash_etal12}
\begin{equation}\label{eq:W}
    \frac{\partial}{\partial \tau}W(\tau,\tau') = \left[\dot{\mathcal{P}}(\tau),\mathcal{P}(\tau)\right] W(\tau,\tau'),\quad W(\tau',\tau')=\mathbb{I}\,,
\end{equation}
where the projection $\mathcal{P}(\tau)$ is defined in Eq.~\eqref{eq:PandQ_superprojectors}, and the dot indicates the derivative with respect to $\tau$. The parallel transport propagator $W$ introduced in Eq.~\eqref{eq:W} is provided in Ref.~\cite{Breuer_Petruccione07} and reads
\begin{equation} \label{eq:propagator_parallel_transport}
    W(\tau,\tau')=\mathcal{T}_{\leftarrow} \exp \left(\int_{\tau'}^\tau ds\, \left[\dot{\mathcal{P}}(s),\mathcal{P}(s)\right]\right),
\end{equation}
with $\mathcal{T}_{\leftarrow}$ the Dyson time-ordering operator and $\left[\dot{\mathcal{P}}(s),\mathcal{P}(s)\right]$ the generator of the evolution. For the qubit system under consideration in this work, the generator in Eq.~\eqref{eq:propagator_parallel_transport} reduces to $\left[\dot{P}_-(\tau),P_-(\tau)\right]$, where $P_-(\tau)$ is an element of the set of projectors $\{P_+,P_-\}\subset\text{ker}(\mathcal{L})$ for the STARE Lindbladian \eqref{eq:dephasingME_two_level}. As shown rigorously in Ref.~\cite{Albash_etal12}, $W(\tau,\tau')P_-(\tau')=P_-(\tau)$. In the next section, the propagator $W(\tau,\tau')$, with the adiabatic generator $\left[\dot{P}_-(\tau),P_-(\tau)\right]$, will enter via the coefficients of the adiabatic power series expansion.

\subsection{Power series expansion}\label{app:ss:expansion}
Using Poincar\'e’s definition of asymptotic power series, it is possible to express the solution of the STARE master equation in terms of a recursive relation that encodes the adiabatic theorem \cite{Teufel03}. The adiabatic trajectory, in fact, can be identified with the dynamics at lowest order in an expansion in the rate of variation of the Liouvillian. Recall that $T$ is the total transfer time, with its inverse $1/T$ loosely interpreted as the average rate of variation of the Hamiltonian. At finite $T$, corrections scale with the small parameter $1/\lambda_0$ where $\lambda_0\propto T$.  The quantity $1/\lambda_0$ plays the role of an adiabaticity parameter. For unitary dynamics, $\lambda_0\ge g_0T$, where $g_0$ is the smallest gap of the instantaneous energy spectrum. In the presence of non-unitary Lindblad dynamics, the adiabatic theorem can then be extended to compare the spectral gap of the Lindbladian \cite{Minganti:2018}, $\sqrt{g_0^2+\gamma^2},$ with the characteristic time of the transfer $T$, such that $\lambda_0\equiv T\sqrt{g_0^2+\gamma^2}\gg 1$ determines the adiabaticity parameter $1/\lambda_0$ for the STARE Lindbladian. 

Now assuming that the Lindbladian spectrum is gapped, as is here the case, we can write the density matrix as \cite{Avron_etal12_CMP}
\begin{equation} \label{eq:expansion_of_rho}
    \varrho(\tau) = \sum_{n=0}^{N} \left[ a_n(\tau)+b_n(\tau)\right] + \lambda_0^{-N-1}r_N(T,\tau)\,,
\end{equation}
where $a_n(\tau)\in\text{ker}(\mathcal{L}(\tau))$ and $b_n(\tau)\in\text{ran}(\mathcal{L}(\tau))$ are the expansion coefficients at $n$-th order in $1/\lambda_0$. These coefficients are determined by the following recurrence relations \cite{Avron_etal12_CMP}
\begin{align}\label{eq:recursive_relation}
    &b_0(\tau) =0\,,\\
    &a_n(\tau) = W(\tau,0)a_n(0)+\int_0^{\tau}d\tau'\, W(\tau,\tau')\dot{\mathcal{P}}(\tau')b_n(\tau')\,,\nonumber\\
&b_{n+1}(\tau) = \mathcal{L}^{-1}(\tau) \dot{\mathcal{P}}(\tau) a_n(\tau)+\mathcal{L}^{-1}(\tau) \mathcal{Q}(\tau) \dot{b}_n(\tau)\,, \nonumber
\end{align}
where $W$ is the parallel transport propagator of Eq.~\eqref{eq:propagator_parallel_transport}, $\mathcal{P}$ and $\mathcal{Q}$ are the superprojectors defined in Eq.~\eqref{eq:PandQ_superprojectors}, and $\mathcal{L}^{-1}$ is the inverse of the STARE Lindbladian, defined on $\text{ran}(\mathcal{L})$. The remainder term $r_N$ in Eq.~\eqref{eq:expansion_of_rho} is determined using Duhamel's principle and is given explicitly in Refs.~\cite{Avron_etal12_CMP,Fraas_Hanggli17}. For our purposes, we will only consider an expansion of the density matrix $\varrho$ to $\mathcal{O}(\lambda_0^{-3})$, such that
\begin{align} \label{eq:density_matrix_expansion_ito_a_b}
    \varrho(\tau) = \,\,&a_0(\tau) + \left[a_1(\tau)+b_1(\tau)\right]\\
    &+\left[a_2(\tau)+b_2(\tau)\right] + \mathcal{O}(\lambda_0^{-3})\,.\nonumber
\end{align}
In Appendix~\ref{app:density_matrix_expansion} we derive the explicit form of the coefficients using relation \eqref{eq:recursive_relation} and obtain:
\begin{eqnarray}\label{eq:list_of_density_matrix_coeff_a_and_b}
a_0(\tau)&=& P_-(\tau)\,,\\  
a_1(\tau)&=&(P_-(\tau)-P_+(\tau)) \mathcal{J}(\tau)\,,\nonumber\\
b_1(\tau)&=&P_+(\tau)\dot{P}_-(\tau)/\lambda_{+}+\dot{P}_-(\tau)P_+(\tau)/\lambda_{-}\,,\nonumber\\
a_2(\tau)&=&(P_-(\tau)-P_+(\tau)) \nonumber\\
& \times&   \int_0^{\tau}d\bar{\tau}\, \text{Tr}\left\{P_+(\bar{\tau})[\dot{P}_-(\bar{\tau})]^2\right\} \big(x_1(\bar{\tau}) + x_2(\bar{\tau})\big),\nonumber\\
b_2(\tau)&=& \left(x_1(\tau) P_+(\tau) \dot{P}_-(\tau) + x_2(\tau)\dot{P}_-(\tau) P_+(\tau)\right)\nonumber,
\end{eqnarray}
where we have introduced 
\begin{equation}
\label{eq:J}
    \mathcal{J}(\tau) = \int_0^\tau d\tau'\frac{2\text{Re}[\lambda_{-}(\tau')]}{|\lambda_{-}(\tau')|^2} \,\text{Tr}\left\{P_+(\tau')[\dot{P}_-(\tau')]^2\right\}  ,
\end{equation}
which is negative, $\mathcal{J}(\tau)\leq 0 $.
Moreover,
\begin{equation}\label{eq:x1_main}
    x_1(\tau) = \left(\frac{2\mathcal{J}(\tau)}{\lambda_{+}(\tau)}-\frac{\dot{\lambda}_{+}(\tau)}{\lambda^3_{+}(\tau)} +\frac{\frac{d}{d\tau}(\leftindex_\tau{}\langle +|\dot{-}\rangle_\tau)}{\lambda^2_{+}(\tau) \leftindex_\tau{}\langle+|\dot{-}\rangle_\tau} \right),\nonumber
\end{equation}
\begin{equation}\label{eq:x2_main}
    x_2(\tau) = \left(\frac{2\mathcal{J}(\tau)}{\lambda_{-}(\tau)}-\frac{\dot{\lambda}_{-}(\tau)}{\lambda^3_{-}(\tau)} +\frac{\frac{d}{d\tau}(\leftindex_\tau{}\langle \dot{-}|+\rangle_\tau)}{\lambda^2_{-}(\tau)\leftindex_\tau{}\langle \dot{-}|+\rangle_\tau} \right).\nonumber
\end{equation} 
In the above expressions the dot indicates a derivative with respect to $\tau$, while
\begin{equation}\label{eq:Lindbladian_eigenvalues_MAIN}
     \lambda_{\pm}(\tau) = [-\gamma(\tau) \pm i  \Delta(\tau) ]T\,
\end{equation}
are (dimensionless) eigenvalues of the STARE Lindbladian \eqref{eq:dephasingME_two_level}. The real part of $\lambda_{\pm}(\tau)$ is determined by the dephasing rate $\gamma(\tau)$, while the imaginary part depends on the gap $\Delta(\tau)=E_+(\tau)-E_-(\tau)$ of the isolated two-level system, see Eq.~\eqref{eq:generic_energies_2Level}. Indeed, \mbox{$\lambda_0=|\lambda_{\pm}(1/2)|$} is the rescaled spectral gap. We remark that the adiabatic power series expansion is also valid for the unitary dynamics. In this case, the Liouvillian is the von-Neumann equation, the eigenvalues $\lambda_{\pm}$ are imaginary and the coefficient $a_1$, being proportional to the real part of $\lambda_{-}$, vanishes identically. 

In the following, we use the expansion \eqref{eq:density_matrix_expansion_ito_a_b} to obtain an analytic form for the infidelity and optimal schedule of the STARE Lindbladian.

\subsection{The infidelity}
\label{Sec:Infidelity}
To assess the efficiency of the adiabatic transfer for the STARE protocol, we again consider the infidelity
\begin{equation} \label{eq:infidelity_trace_definition}
\mathcal{I} = 1-\text{Tr}\{P_-(1)\varrho(1)\}\,.
\end{equation}
Upon inserting $\varrho(1)$ from the expansion in Eq.~\eqref{eq:density_matrix_expansion_ito_a_b} we obtain, to order $1/\lambda_0^2$ in the adiabaticity parameter,
\begin{equation} \label{eq:infidelity_analytic_Avron}
	\mathcal{I}(a,b,d_i,d_f) = 2 \int_0^1 d\tau\, M(q)\left(\frac{dq}{d\tau}\right)^2 + \mathcal{C}\,,
\end{equation}
where, as shown in Appendix~\ref{app:Avron_derivation},
\begin{equation}\label{eq:M_function}
	M(q) = \frac{b  (d_f-d_i)^2}{4 \left(d(q)^2+1\right)^2 \left(a^2 \left(d(q)^2+1\right)+b^2\right)}\geq 0\,.
\end{equation}
On the right-hand side of the expression above, the sweep function $d$, Eq.~\eqref{eq:sweep_function_dimensionless_parameters}, and the dephasing rate $b$ of Eq.~\eqref{eq:dimensionless_parameters_b} are functions of the schedule $q=q(\tau)$. The expression for the infidelity \eqref{eq:infidelity_analytic_Avron} holds when the instantaneous spectrum is gapped at every instant of time. This is the case for the dynamics we consider, where the gap is bound from below by $g_0$. The leading-order term in Eq.~\eqref{eq:infidelity_analytic_Avron} was first derived, also in a more general setting, in Ref.~\cite{Avron_etal10}.

The quadratic-order contribution to the infidelity~\eqref{eq:infidelity_analytic_Avron} is contained in $\mathcal{C}$, which takes the form
\begin{align} \label{eq:C_q_function}
    &\mathcal{C}(a,b,d_i,d_f) = 2 \int_0^1 d\tau\,  \dot{q}^2 M(q) \Bigg[2\mathcal{J}(\tau) - d(d_f-d_i)   \nonumber\\
    & \, \left.  \times  \dot{q} \frac{ \left(3 a^4 \left(d^2+1\right)^2-3 a^2 b^2 \left(d^2+1\right)-2 b^4\right)}{b \left(d^2+1\right) \left(a^2 \left(d^2+1\right)+b^2\right)^2}  \right]
\end{align}
when $b$, and therefore the dephasing rate $\gamma$, are constant. 

For unitary dynamics, $b=0$, the terms containing $\mathcal J (\tau)$ are identically zero, and the leading-order contribution to the infidelity \eqref{eq:infidelity_analytic_Avron} is solely given by $\mathcal{C}$. In this case, the infidelity $\mathcal{I}\sim \dot{q}^3/(a^2 (d^2+1)^2)$ scales with the time derivative of the Hamiltonian schedule, and the optimal protocol can be found following the treatment of Ref.~\cite{Roland_Cerf02}. Focusing on the opposite limit, in which $b\gg a$, the dominant contribution comes from the leading-order term of Eq.~\eqref{eq:infidelity_analytic_Avron}, and scales as $\mathcal{I} \sim \dot{q}^2/b$ for constant dephasing rates. Evidently, the infidelity strongly depends on the interplay between the parameters $a$ and $b$. 

In what follows, we consider the regime where $b\gg a^2$. For the linear schedule, for instance, this corresponds to $\gamma T\gg \epsilon^4 T^6/g_0^2$ and establishes the strictest lower bound on the dephasing rate $b$ to ensure the dominance of the first term in Eq.~\eqref{eq:infidelity_analytic_Avron}. Crucially, this permits the treatment of $\mathcal{C}$ as a correction term, as we will show in the next section.

\subsection{Optimal schedule} \label{ss:optimal_sweep_fucntion_derivation}
The infidelity in Eq.~\eqref{eq:infidelity_analytic_Avron} can be regarded as a functional of the schedule $q(\tau)$, which controls how both the Hamiltonian and projectors in the STARE master equation \eqref{eq:Lindbladian_dimensionless} vary in time. In this section we search for a choice of this schedule which minimizes the infidelity. We neglect the correction $\mathcal{C}$ in this minimization procedure, since it will be seen to make a negligible contribution to the infidelity in the regime where $b\gg a^2$. 

\begin{figure}[t]
    \centering
    \includegraphics[width=\columnwidth]{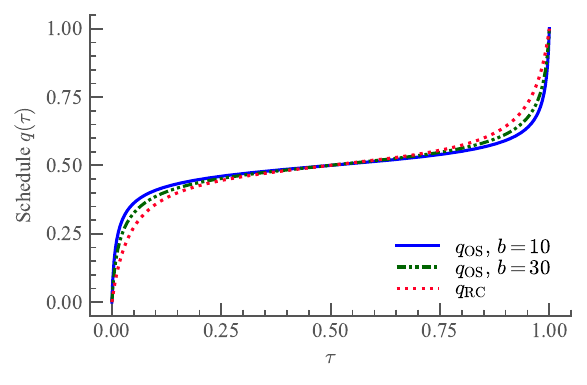}
    \caption{The optimal schedule $q_{\rm OS}(\tau)$ of Eq.~\eqref{eq:Avron_schedule_dimensionless} as a function of the dimensionless time $\tau$. Results are shown for $b=10,\,30$, which scales the incoherent dynamics. For reference, the schedule $q_{\rm RC}$ \eqref{eq:RC_schedule_dimensionless} of unitary dynamics is shown. All curves have been calculated taking a symmetric sweep with $d_f=-d_i=8$ and $a=10$.}
    \label{fig:scheduleOS}
\end{figure}

The leading-order term in Eq.~\eqref{eq:infidelity_analytic_Avron} can be viewed as an action functional with Lagrangian $L(\tau,q(\tau),\dot{q}(\tau))=2M(q)\dot{q}^2$ and a $q$-dependent `mass' function $M(q)$ \cite{Avron_etal10}. Importantly, this Lagrangian does not depend on $\tau$ explicitly, and so its value remains constant along a minimizing orbit. The evolution schedule $q_\text{OS}$ minimizing the functional is then determined from the optimal speed along the minimizing path \cite{Avron_etal10}
\begin{equation} \label{eq:optimal_speed_Avron_schedule}
	\frac{dq_\text{OS}}{d\tau} = \sqrt{\frac{\xi}{M(q_\text{OS})}}\,,
\end{equation}
where the constant $\xi$ is to be fixed by the boundary conditions $q_\text{OS}(0)=0$ and $q_\text{OS}(1)=1$. From its form in Eq.~\eqref{eq:M_function}, $M(q)$ is seen to be a decreasing function of $d(q)^2$, and is therefore maximal at the avoided level-crossing where $d=0$. See also Eqs.~\eqref{eq:HqGeneral} and \eqref{eq:sweep_function_dimensionless_parameters}. Integrating Eq.~\eqref{eq:optimal_speed_Avron_schedule} leads to expressions for $\xi$ and $q_\text{OS}(\tau)$ in terms of 
\begin{equation}\label{eq:definition_of_x_in_infidelity_expr}
	x(d) \equiv  \frac{b d}{\sqrt{a^2 \left(d^2+1\right)+b^2}}\,,
\end{equation}
and
\begin{equation}
	\theta_i=\arctan x(d_i)\quad{\rm and}\quad\theta_f=\arctan x(d_f)\,.
\end{equation}
We find that $\xi = (\theta_f-\theta_i)^2/(4\gamma)$, while the optimal schedule reads
\begin{align}\label{eq:Avron_schedule_dimensionless}
    q_\text{OS}(\tau) &=  \frac{(a^2+b^2)^{1/2}\tan \left[\tau (\theta_f-\theta_i) + \theta_i\right]}{(d_f-d_i)\left(b^2-a^2 \tan ^2\left[\tau (\theta_f-\theta_i) + \theta_i\right]\right)^{1/2}} \nonumber\\
    &\quad -\frac{d_i}{d_f-d_i}\,.
\end{align}
Interestingly, for $b\gg a$, the schedule $q_\text{OS}$ exhibits the same functional dependence on time as the optimized schedule $q_\mathrm{RC}$ for purely coherent dynamics, Eq.~\eqref{eq:RC_schedule_dimensionless}, see Fig.~\ref{fig:scheduleOS}. The lowest attainable infidelity $\mathcal{I}_{\min}$ for $b\gg a^2$ is found by inserting Eq.~\eqref{eq:Avron_schedule_dimensionless} into Eq.~\eqref{eq:infidelity_analytic_Avron}. After integrating, we obtain
\begin{equation}\label{eq:Imin}
    \mathcal{I}_{\min} =  \frac{1}{2 b} (\theta_f-\theta_i)^2\,,
\end{equation}
which is valid up to order $\mathcal{O}(\lambda_0^{-1})$. Assuming $|d_f|,|d_i|\gg 1$, the minimum infidelity of Eq.~\eqref{eq:Imin} reduces to the form
\begin{equation}
    \mathcal{I}_{\min} \approx \frac{2\arctan^2\left(b/a\right)}{b}\,.
\end{equation}
For $b\gg a$, corresponding to $\gamma\gg g_0$, the leading-order contribution to the infidelity therefore scales with the inverse of the dephasing rate as $\mathcal{I}_{\min}\approx \pi^2/(2b)$.

We also computed the first higher-order correction $\mathcal{C}$ \eqref{eq:C_q_function} for $q=q_\text{OS}$: 
     \begin{align}\label{eq:Cq_for_optimized_schedule}
        \mathcal{C} = -\mathcal{I}_{\min}^2 &- \frac{\mathcal{I}_{\min}}{b}  \left[5 \ln \left(\frac{d_f x(d_i)}{d_i x(d_f)}\right) - \ln \left(\frac{d_f^2+1}{d_i^2+1}\right) \right.\nonumber \\
        &\left. 
        + 2 \left(\frac{x^2(d_f)}{d_f^2}-\frac{x^2(d_i)}{d_i^2}\right) \right].
    \end{align}
Note that the second term of Eq.~\eqref{eq:Cq_for_optimized_schedule} is zero for a symmetric sweep function, for which $\mathcal{C}= -\mathcal{I}_{\min}^2$. Evidently, in the parameter regime of interest, where $\mathcal{I}_{\min}\ll1$, the contribution of this correction to the infidelity is negligible.

\section{Efficiency of the noise-assisted protocols}
\label{Sec:efficiency}
Now we analyze the efficiency of adiabatic transfer based on the STARE Lindbladian of Eq.~\eqref{eq:dephasingME_two_level}, and investigate its potentially advantageous impact for reducing the infidelity or, equivalently, speeding up the transfer. We then perform a comparison with the optimized coherent protocol implemented under ideal conditions, namely, in the absence of noise and dissipation. From this comparison, we identify the parameter regime in which the incoherent dynamics governed by the STARE Lindbladian outperforms the optimal coherent protocol of Ref.~\cite{Roland_Cerf02}. 

\begin{figure}[t]
    \centering
    \includegraphics[width=\columnwidth]{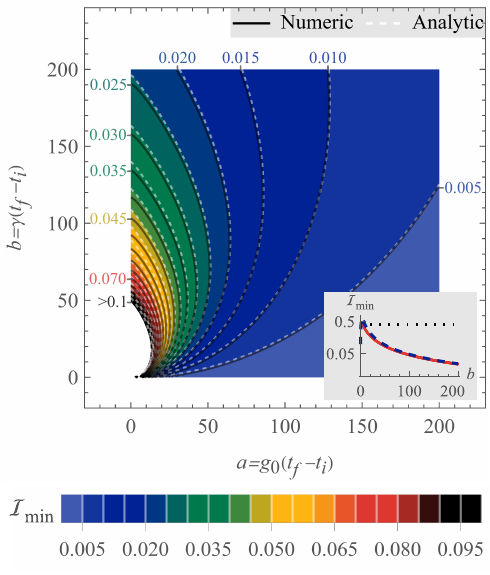}
     \caption{The minimum attainable infidelity $\mathcal{I}_{\min}$ \eqref{eq:Imin} as a function of the dimensionless adiabaticity parameter $a$ and the dephasing strength $b$. Solid contours are obtained by numerically integrating the STARE master equation~\eqref{eq:Lindbladian_dimensionless} using the optimal schedule in Eq.~\eqref{eq:Avron_schedule_dimensionless}. The dashed contours display the analytic results, given by Eq.~\eqref{eq:Imin}. For reference, we indicate the numeric value of the infidelity along several contours, with the white region in the lower left corner corresponding to $\mathcal{I}_{\min}>0.1$. The inset shows the minimum infidelity of the STARE protocol as a function of $b$ and for $a=2$. The analytic result (dashed) approximates the exact numeric result (solid) well. The dotted line indicates the minimal infidelity achieved by the optimal unitary dynamics for the same transfer time. We set $d_f=-d_i=100$.}
    \label{fig:Infid_numeric_and_analytic_results}
\end{figure}

\begin{figure}[t]
    \centering
    \includegraphics[width=1.035\columnwidth]{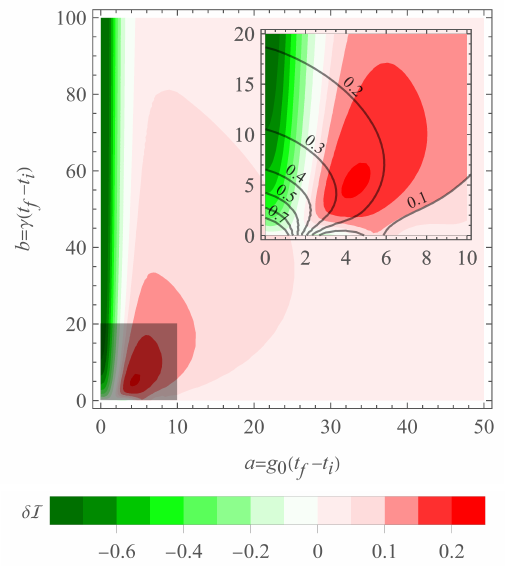}\vspace{0.11cm}
    \caption{Contour plot of the difference $\delta\mathcal{I}=\mathcal{I}_\mathrm{eng}-\mathcal{I}_\mathrm{uni}$ between the minimum infidelity $\mathcal{I}_{\mathrm{uni}}$ attained at the end of the transfer time using the unitary dynamics with the optimal schedule, Eq.~\eqref{eq:RC_schedule_dimensionless}, and the infidelity $\mathcal{I}_\mathrm{eng}$ achieved with the STARE Liouvillian \eqref{eq:dephasingME_two_level} using the STARE schedule of Eq.~\eqref{eq:Avron_schedule_dimensionless}. The difference is shown in the $a$--$b$ plane. The unitary dynamics outperforms the STARE dynamics in the adiabatic regime. Meanwhile, negative values of $\delta\mathcal{I}$, observed for non-adiabatic dynamics, signify better state transfer when implementing the STARE protocol. The inset shows the region where \mbox{$a\leq10$} and $b\leq20$ in more detail. Contours provide the exact numerical values of $\mathcal{I}_\mathrm{eng}$ in the $a$--$b$ plane; see Fig.~\ref{fig:Infid_numeric_and_analytic_results}. Here, $d_f=-d_i=10$.}
    \label{fig:comparison}
\end{figure}

\subsection{Infidelity of adiabatic transfer}
We assess the efficiency of the STARE protocol by comparing the infidelity for the optimal schedule $q_\text{OS}(\tau)$ \eqref{eq:Avron_schedule_dimensionless} with the one of the unitary dynamics using the optimal schedule $q_\mathrm{RC}(\tau)$ \eqref{eq:RC_schedule_dimensionless}. For these protocols, the infidelity of the solutions to the master equation \eqref{eq:Lindbladian_dimensionless} are naturally functions of the dimensionless parameters $a$ and $b$, Eqs.~\eqref{eq:dimensionless_parameters_a} and \eqref{eq:dimensionless_parameters_b}, as well as $d_i$ and $d_f$ from Eq.~\eqref{eq:HqGeneral}. In particular, $a \sim 1$ separates the adiabatic from the non-adiabatic regime of the coherent dynamics. The variable $b$, in turn, quantifies the strength of the dephasing processes induced by the STARE Lindbladian, such that for $b\gg 1$ they become significant. We choose $d_i$, $d_f$ such that $|d_f|,|d_i|\gg 1$. In fact, it is found that already for $|d_f|,|d_i|\gtrsim 10$ the infidelity $\mathcal{I}$ has essentially reached its limiting value. For symmetric schedules, we therefore conveniently study the infidelity \eqref{eq:Imin} as a function solely of two variables, $\mathcal{I}_{\min}(a,b)$.

Figure \ref{fig:Infid_numeric_and_analytic_results} displays a contour plot of the infidelity of the STARE protocol as a function of the adiabaticity and dephasing parameters, $a$ and $b$, at fixed transfer time $T$. Horizontal slices through the figure depict the monotonically decreasing behavior of the infidelity with increasing $a$, confirming that slower evolution (being `more adiabatic') has the tendency to suppress transitions of the system out of its ground state. The behavior of $\mathcal{I}_{\min}$ with $b$ depends non-trivially on $a$. At small $a$ but $b\gg 1$, the infidelity is inversely proportional to the dephasing rate, $\mathcal{I}_{\min}\sim 1/b$.

In Fig.~\ref{fig:comparison} we assess when the STARE protocol with its optimal schedule, Eq.~\eqref{eq:Avron_schedule_dimensionless}, outperforms the optimal coherent protocol, Eq.~\eqref{eq:RC_schedule_dimensionless}, implemented under ideal conditions. For this purpose, we display the difference between the corresponding infidelities, $\delta\mathcal{I}=\mathcal{I}_\mathrm{eng}-\mathcal{I}_\mathrm{uni}$, where $\mathcal{I}_\mathrm{uni}$ and $\mathcal{I}_\mathrm{eng}$ are the infidelities achieved for the purely coherent dynamics and STARE Lindbladian, respectively. A positive value here corresponds to a lower infidelity (higher efficiency) of the unitary adiabatic transfer with the optimal protocol by Roland and Cerf~\cite{Roland_Cerf02}, Eq.\ \eqref{eq:RC_schedule_dimensionless}. A negative value indicates that the STARE adiabatic transfer is more efficient. This occurs at small values of $a$, in the non-adiabatic regime, and for large dephasing rates $b$. Note that this region corresponds to an area satisfying $b\gg a^2$, see Sec.\ \ref{Sec:Infidelity}.

We close this section with a discussion on the schedule and infidelity in the two most representative limiting cases. When the dephasing rate is very weak, with $b\rightarrow0$, we find that the optimal schedule is that of the purely unitary dynamics~\eqref{eq:RC_schedule_dimensionless}. This schedule will lead to the smallest possible infidelities in a fixed transfer time $T$ \cite{Roland_Cerf02}. The infidelity then decreases with the transfer time, that is fixed by the adiabaticity parameter $a$. In the opposite limit, $b\rightarrow\infty$, the dynamics is governed by the dissipator term $\mathcal{D}(t)\varrho(t)$ of the STARE Lindbladian \eqref{eq:dephasingME_two_level}. The strong dephasing mechanism results in the suppression of transitions out of one of the branches, thereby stabilizing the target trajectory. This dynamics is reminiscent of the quantum Zeno effect \cite{Menu_etal22,Paz-Silva_etal12,Raimond_etal12}. For this case, it is best to use the optimal schedule of Eq.~\eqref{eq:Avron_schedule_dimensionless} that is tailored for open-system dynamics. 
Both schedules maximize the time spent in the region where the dynamics crosses the minimal gap. In the limit $b\gg a^2$, however, the characteristic timescale is set by the dephasing rate $\gamma$. In the next section we discuss the implications on the optimal transfer time.

\subsection{Optimal transfer time} \label{ss:analytic_results_I_and_T}
Next, we discuss the minimum transfer time that the STARE protocol can achieve. In fact, from Eq.~\eqref{eq:Imin} it is possible to derive an expression for the minimum transfer time $\mathcal{T}_{\min}$ as a function of the infidelity $\mathcal{I}$. This task is simplified by noting that the dependence on $T$ on the right of $x(d)$ in Eq.~\eqref{eq:definition_of_x_in_infidelity_expr} cancels, rendering $x(d)$ a function only of $d$ and $\gamma/g_0$. The minimum transfer time is then found to be
\begin{equation}\label{eq:Tmin}
    \mathcal{T}_{\min} = \frac{2\xi}{\mathcal{I}}=\frac{1}{2 \gamma \mathcal{I}} \left[ \arctan x(d_f)-\arctan x(d_i) \right]^2. 
\end{equation}
As expected, achieving low infidelities comes at the expense of longer transfer times. Interestingly, at strong dephasing, $\mathcal{T}_{\min}$ scales with the inverse of the dephasing rate. By way of illustration, examine the behavior of the infidelity $\mathcal{I}_{\min}$ along a vertical axis in Fig.~\ref{fig:Infid_numeric_and_analytic_results} where $a<50$. This seems to suggest that, by increasing the dephasing rate $\gamma$, faster transfers can be achieved. This is in contrast with the scaling of the transfer time when implementing the protocol by Roland and Cerf \cite{Roland_Cerf02}, which is bound by the inverse value of the gap $1/g_0$. Thus, for values $\gamma\gg g_0$ (corresponding to $b\gg a$), the protocol based on engineered dephasing predicts a shortcut to adiabaticity that can significantly exceed the efficiency of the optimal, unitary dynamics.  In Sec.~\ref{s:speed_limit} we unravel this  behavior with a detailed study of the regimes of validity of the STARE Lindblad master equation~\eqref{eq:Lindbladian_dimensionless} and specifically of the associated speed limit. 

\section{Speed limit} \label{s:speed_limit}
In the previous section we showed that, in the strong dephasing regime, the lower bound on the transfer time of the STARE protocol, ${\mathcal T}_{\rm min}$ \eqref{eq:Tmin}, is set by the inverse of the dephasing rate. This rate scales a time-dependent Lindbladian, whose jump operators are the projectors onto the instantaneous eigenstate of the Hamiltonian, see Eq.~\eqref{eq:dephasingME_two_level}. Therefore, the requirement $\gamma\gg g_0$ for observing a significant shortcut to adiabaticity with respect to the optimal unitary dynamics makes several demands of the underlying physical model. The extent to which these can be realized will set a stricter lower bound on ${\mathcal T}_{\rm min}$. Below we argue that the physical model restricts the validity of the master equation to values for which $\gamma\ll g_0$, showing that the minimal gap of the Hamiltonian ultimately limits the speed of the STARE protocol.

\begin{figure}[t]
    \centering
    \includegraphics[width=\columnwidth]{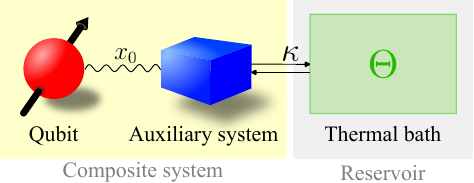}
    \caption{A possible physical realization of the STARE Lindbladian. The composite system \eqref{eq:interaction_Hamiltonian_for_ME_general} consists of a qubit and an auxiliary system, where $x_0$ scales the coupling. The auxiliary system, in turn, is coupled to a thermal bath at temperature $\Theta$ and thermalizes at rate $\kappa$. The dynamics of the composite system is described by master equation~\eqref{eq:MEq}.}
    \label{fig:qubit-aux}
\end{figure}

\subsection{Constructing the STARE Lindbladian}\label{ss:intro to microscopic model}
The derivation of the STARE Lindbladian of Eq.~\eqref{eq:dephasingME_two_level} from a microscopic model has been shown in Ref.~\cite{Menu_etal22}. The underlying physical model is that of a qubit coupled to a damped auxiliary system, shown schematically in Fig.~\ref{fig:qubit-aux}. Here we review the basic steps in order to determine the bounds on the dephasing rate that the Lindbladian's regime of validity imposes. We describe the dynamics of the density matrix $\chi$ on the composite Hilbert space of the qubit and auxiliary system by
\begin{equation}
\label{eq:MEq}
\partial_t\chi=-i[H_\mathrm{SA}(t) ,\chi(t)] + \kappa \mathcal{L}_\mathrm{A}(\chi(t))\,,
\end{equation}
where $H_\mathrm{SA}(t)$ is the Hamiltonian of the composite system and the superoperator $\mathcal{L}_\mathrm{A}(\chi(t))$ is a Lindbladian whose jump operators act in the Hilbert space of the auxiliary system $\mathcal H_A$. We impose that $[H_\mathrm{SA}(t),H_\mathrm{S}(t)\otimes  \mathbb{I}_\mathrm{A} ]=0$, where $H_{\mathrm{S}}$ is the qubit Hamiltonian \eqref{eq:LZ_Hamiltonian}, and assume that the Lindbladian $\mathcal{L}_\mathrm{A}$ describes the action of a thermal bath at temperature $\Theta$. We denote by $\kappa$ the rate of thermalization. In Ref.~\cite{Menu_etal22}, it was shown that master equation \eqref{eq:MEq} can be reduced to the STARE Lindbladian \eqref{eq:dephasingME_two_level} by taking
\begin{equation}\label{eq:interaction_Hamiltonian_for_ME_general}
    H_\mathrm{SA}(t) = H_\mathrm{S}(t)\otimes \mathbb{I}_\mathrm{A} + \mathbb{I}_\mathrm{S}\otimes H_\mathrm{A} +  x_0 H_\mathrm{S}(t)\otimes X_\mathrm{A}\,, 
\end{equation}
where $H_\mathrm{A}$ is the Hamiltonian of the auxiliary system in the absence of coupling, and operator $X_\mathrm{A}$ is defined in the Hilbert space $\mathcal H_A$ such that $[X_\mathrm{A},H_\mathrm{A}]\neq 0$. The parameter $x_0$ is  dimensionless and scales the coupling. See Fig.~\ref{fig:qubit-aux} for an illustration. The STARE Lindbladian is derived by extending the framework of the adiabatic master equation of Ref.~\cite{Albash_etal12} to a dynamics where the coupling between qubit and auxiliary system is time-dependent and varies with the same schedule as the system Hamiltonian $H_{\mathrm{S}}$. In the weak-coupling limit, the resulting Lindbladian has the form of Eq.~\eqref{eq:dephasingME_two_level}, with the dephasing strength given by \cite{Menu_etal22}
\begin{equation}\label{eq:gamma_t}
    \gamma(t) = \frac{1}{2} \Gamma_\mathrm{re}(0)[E_+(t)-E_-(t)]^2\,,
\end{equation}
where $E_\pm$ are the qubit Hamiltonian eigenenergies of Eq.~\eqref{eq:generic_energies_2Level}. The rate is now time-dependent, and scales with the square of the instantaneous gap. The proportionality factor $\Gamma_\mathrm{re}(0)$ is the Fourier component at $\omega=0$ of the Fourier transform of the auxiliary system's autocorrelation function, 
\begin{align}\label{eq:spectral_density_Re_MAIN}
     \Gamma_\mathrm{re}(\omega) &= \int_{-\infty}^\infty d\tau\, e^{i\omega\tau} C_\mathrm{A}(\tau)\,,\\
     C_\mathrm{A}(\tau) &\equiv x_0^2\,\text{Tr}\left\{U^\dagger(\tau){X}_\mathrm{A}U(\tau){X}_\mathrm{A}\varrho_\mathrm{A} \right\},\nonumber
\end{align}
with $U(\tau)=\exp(-iH_\mathrm{A} \tau)$ and $$\varrho_\mathrm{A}=\exp(-H_\mathrm{A}/(k_B\Theta))/\text{Tr}\left\{\exp(-H_\mathrm{A}/(k_B\Theta)\right\}$$
the density matrix of the auxiliary system in the Born-Markov limit.
Details of the derivation are reviewed in Appendix~\ref{app:microscopic_derivation}. 

\begin{figure*}[t]
    \centering
    \includegraphics[width=0.685\columnwidth]{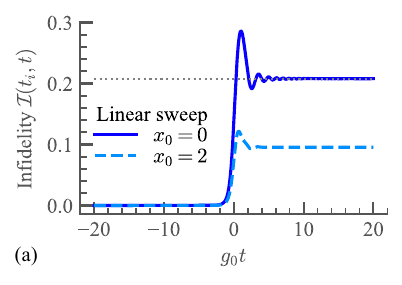}
    \includegraphics[width=0.685\columnwidth]{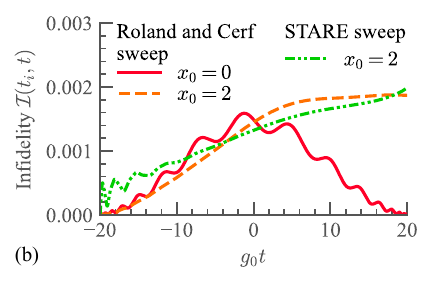}
    \includegraphics[width=0.685\columnwidth]{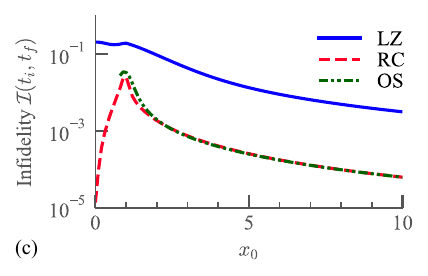}
    \caption{Subplots (a) and (b) display the infidelity $\mathcal{I}$, Eq.~\eqref{eq:tunnelingProb_LZ_general}, as a function of the rescaled time $g_0t$ obtained by integrating Eq.~\eqref{eq:MEq} for the qubit coupled to a damped auxiliary two-level system. In (a) the sweep is linear, with the solid line corresponding to the unitary case ($x_0=0$) and the dashed line to the dissipative case with $x_0=2$. In (b) the corresponding curves are shown for the sweep by Roland and Cerf \cite{Roland_Cerf02}. For comparison, we also provide the infidelity for $x_0=2$ using the schedule of the STARE Lindbladian, Eq.~\eqref{eq:Avron_schedule_dimensionless}. Subplot (c) summarizes the results for the final infidelity $\mathcal{I}$ as a function of the coupling parameter $x_0$. Results are shown for three schedules: the linear schedule (LZ, solid), the Roland and Cerf schedule of Eq.~\eqref{eq:RC_schedule_dimensionless} (RC, dashed), and the STARE schedule of Eq.~\eqref{eq:Avron_schedule_dimensionless} (OS, dot-dashed), with $x_0\gtrapprox0.8$ in the latter case to ensure that the requirement $T^{-1}\ll \gamma_0$ is met. Remaining parameters are set to $\omega_a=\kappa=g_0$, $t_f = -t_i=20 / g_0$ and $n=0$. }
    \label{fig:LZRC_dynamicswithancilla}
\end{figure*}

\subsection{Conditions of validity}
The quantities introduced in Sec.~\ref{ss:intro to microscopic model} are essential for assessing the validity of the approximation that enables us to derive the STARE Lindbladian of Eq.~\eqref{eq:dephasingME_two_level} from Eq.~\eqref{eq:MEq}. An important parameter is the  timescale $\tau_\mathrm{A}$ characterizing the dynamics of the autocorrelation function $ C_\mathrm{A}(\tau) $ of the auxiliary system, and thus the frequency width of its spectrum $\Gamma_\text{re}(\omega)$. We first ignore the temporal variation of the qubit's Hamiltonian. In the weak coupling limit, $x_0\ll 1$, the timescale $\tau_\mathrm{A}$ is determined by the thermalization rate $\kappa$, namely $\tau_\mathrm{A}\sim 1/\kappa$. In order to neglect memory effects due to the coupling with the auxiliary system, corresponding to the Markov approximation, the characteristic timescale $\tau_\mathrm{S}$ over which the system evolves must satisfy $\tau_\mathrm{S}\gg\tau_\mathrm{A}$. For a time-independent Hamiltonian, $\tau_\mathrm{S}\sim 1/\gamma$. Taking the minimum value $\gamma(0)$, we obtain the inequality $\gamma\tau_\mathrm{A}\ll 1$. Using that $C_\mathrm{A}(\tau)\sim  x_0^2 \,e^{-\tau/\tau_\mathrm{A}}$ in our model, then from Eq.~\eqref{eq:gamma_t} we have $\gamma\sim x_0^2g_0^2\tau_\mathrm{A}$, and the timescale separation condition requires 
\begin{equation}\label{eq:validity_condition_1}
(x_0g_0\tau_\mathrm{A})^2\ll 1\,.
\end{equation}
Additionally, for the STARE Lindbladian, changes in the system's instantaneous energy eigenbasis $\vert\pm\rangle_t$, see Eq.~\eqref{eq:generic_eigenstates_2Level}, will be negligible on the timescale of the auxiliary system $\tau_\mathrm{A}$, demanding that \cite{Albash_etal12}
\begin{equation}\label{eq:validity_sec_adiabatic_thm_general}
   \frac{|\leftindex_t{}\langle+|\partial_t{H}_\text{S}(t)|-\rangle_t|}{|E_+(t)-E_-(t)|}\tau_\mathrm{A}\ll 1
\end{equation}
is satisfied at each time $t$. The most stringent condition is at the time instant $t=0$ where the energy gap $|E_+(t)-E_-(t)|$ is minimal and, up to a constant, equal to $g_0$. We compute the transition amplitude using 
Eqs.~\eqref{eq:generic_eigenstates_2Level} and \eqref{eq:LZ_Hamiltonian} with \mbox{$\sin(\theta)=1/\sqrt{1+[s(t)/g_0]^2}$}. Then, it follows that \mbox{$|\leftindex_t{}\langle+|\partial_t{H}_\text{S}(0)|-\rangle_t|\sim \left|\dot{s}(0) \left(1+[s(0)/g_0]^2\right)^{-1/2} \right|$}. Therefore, in order to realize the dynamics of Eq.~\eqref{eq:dephasingME_two_level}, the parameter values for master equation \eqref{eq:MEq} must be chosen such that they comply with the conditions in Eqs.~\eqref{eq:validity_condition_1} and \eqref{eq:validity_sec_adiabatic_thm_general} above.

To understand the implication on the optimal transfer time $\mathcal{T}_{\min}$, see Eq.~\eqref{eq:Tmin}, it is convenient to discuss the regimes of validity in terms of the dimensionless parameters $a=g_0T$ \eqref{eq:dimensionless_parameters_a} and $b=\gamma T$ \eqref{eq:dimensionless_parameters_b}. We use the scaling of the timescale $\tau_\mathrm{A}\sim \gamma/(x_0g_0)^2$ and recast the two inequalities in the form
\begin{eqnarray}\label{eq:regimeOfValidity_1}
&&\frac{b}{a}\ll x_0\,,\\
&&\frac{b}{a}\ll \frac{x_0^2 a}{|\dot d_0|}\,,
\label{eq:regimeOfValidity_2}
\end{eqnarray}
with $d_0=d(\tau_0)$ and the rescaled time $\tau_0=1/2$ corresponding to $t=0$, see Eq.~\eqref{eq:dimensionless_time}.  Keeping in mind the weak coupling condition $x_0\ll1$, these inequalities indicate that the dynamics of master equation~\eqref{eq:dephasingME_two_level} is found from the setup of Fig.~\ref{fig:qubit-aux} for $\gamma\ll g_0$. Hence, in satisfying conditions~\eqref{eq:regimeOfValidity_1} and \eqref{eq:regimeOfValidity_2}, the dynamics is restricted to the regime where the unitary protocol is most efficient and where the transfer time is bounded by $1/g_0$. This result is in agreement with the study in Ref.\ \cite{Albert_etal16}, and shows that the energy gap of the Hamiltonian ultimately determines the speed limit of the STARE protocol of Eq.~\eqref{eq:dephasingME_two_level}.

\subsection{Beyond the Born-Markov limit}\label{ss:beyond_BM}
The derivation of the STARE Lindbladian \eqref{eq:dephasingME_two_level} from a physical model shows that the protocol, as described by Eq.~\eqref{eq:dephasingME_two_level}, can only be realized in the parameter regime where the unitary protocol is more efficient. Therefore, there is no advantage in using the reservoir-engineered STARE protocol over the optimal unitary protocol. Nevertheless, numerical studies of Eq.~\eqref{eq:MEq} with the linear schedule show a definite improvement of the transfer fidelity outside the Born-Markov regime \cite{Menu_etal22}. Specifically, these studies reported a monotonic decrease of the infidelity as the strength of  the coupling $x_0$ is increased, see for example Fig.~\ref{fig:LZRC_dynamicswithancilla}(a). In this section we numerically study the transfer efficiency as a function of $x_0$ for different schedules, including the optimal schedule for the coherent dynamics, Eq.~\eqref{eq:RC_schedule_dimensionless}, and the one of the STARE Lindbladian, Eq.~\eqref{eq:Avron_schedule_dimensionless}. The source code developed is open-source and available online~\cite{king_2024_11058607}.

For this study, we assume that the auxiliary system is a second qubit, with $H_\mathrm{A}=\omega_a\hat{\sigma}_z/2$ and $X_\mathrm{A}=\hat{\sigma}_x$, which in turn is coupled to a bosonic thermal reservoir at temperature $\Theta$. The damping processes of the auxiliary qubit are described by the Lindbladian
\begin{align}
     \mathcal{L}_\mathrm{A}(\chi(t)) &= (n+1)\mathcal{D}(o)\chi + n\mathcal{D}(o^\dag)\chi\,,\nonumber\\
     \mathcal{D}(o)\chi &= o\chi o^\dag - \{o^\dag o,\chi\}/2\,,
\end{align}
where the jump operators $o=\mathbb{I}_\mathrm{S}\otimes\hat{\sigma}_-$ and $o^\dag=\mathbb{I}_\mathrm{S}\otimes\hat{\sigma}_+$ act on the composite Hilbert space $\mathcal{H}_\mathrm{S}\otimes\mathcal{H}_A$, and $n=(\exp[\omega_a/(k_B\Theta)]-1)^{-1}$ is the mean photon number of the harmonic oscillator at frequency $\omega_a$ of the bath in thermal equilibrium at temperature $\Theta$. The dephasing rate $\gamma(t)$ in the Lindblad, Born-Markov limit takes the form given in Eq.~\eqref{eq:gamma_t}, with the qubit autocorrelation function
\begin{equation}
    \Gamma_{\rm re}(\omega=0) = x_0^2 \frac{\kappa(2n+1)}{\kappa^2(2n+1)^2/4+\omega_a^2}\,.
\end{equation}
In Fig.~\ref{fig:LZRC_dynamicswithancilla}(b) we compare the time evolution of the infidelity for the schedule of Roland and Cerf, Eq.~\eqref{eq:RC_schedule_dimensionless}, with that of the optimal schedule of the STARE Lindbladian, Eq.~\eqref{eq:Avron_schedule_dimensionless}, for different values of $x_0$. We note that, tuning $x_0$ with respect to $\kappa$, is equivalent to tuning the memory time of the qubit's effective reservoir, see Ref.\ \cite{Menu_etal22}. Figure~\ref{fig:LZRC_dynamicswithancilla}(c) captures the behavior of the infidelity at the end of the transfer for a range of $x_0$ values. For $x_0>1$ the transfer efficiency  improves by increasing the coupling strength. While an improvement is observed for all protocols, the optimized schedules for the coherent dynamics and engineered dephasing outperform the linear schedule by as much as two orders of magnitude. In general, this result shows the relevance of non-Markovian reservoirs for noise-assisted quantum protocols. 


\section{Summary and Conclusions} \label{s:summary}
 Quantum reservoir engineering is being discussed as an alternative strategy for quantum technologies. The striking advantage is the robustness that these types of protocols offer with respect to protocols based on unitary dynamics. One key question is what is their efficiency with respect to the corresponding optimized protocols for unitary dynamics. In this paper, we have performed a systematic analysis comparing the efficiency of the optimal protocol for unitary adiabatic quantum state transfer in a qubit with that of a protocol based on quantum reservoir engineering. This protocol is based on a Lindbladian which we dubbed by STARE Lindbladian. 

An optimal \textit{open-system} evolution schedule was derived, allowing for the optimization of the STARE-based dynamics. From this optimal schedule, we could derive the lower bound to the transfer time of the STARE Lindblad protocol. Our result rigorously shows that the lower bound to the STARE protocol timescale is determined by the interplay of the energy gap and of the rate of the incoherent dynamics. If it were possible to choose arbitrarily the rate $\gamma$ scaling the STARE Lindbladian, the STARE adiabatic transfer time would be bound by $1/\gamma$. Considering that quantum state transfer is paradigmatic for the adiabatic quantum search, this would provide a new complexity class for STARE protocols. This conclusion is based on postulating the possibility of freely tuning the rate $\gamma$ scaling the STARE Lindbladian. 

A derivation of the STARE Lindbladian from first principles, however, shows that the requirements for the validity of the STARE Lindbladian limit the choice of $\gamma$ to values that are smaller than the energy gap. This result indicates that, for the microscopic model we considered, the STARE protocol provides no net advantage with respect to the optimal unitary protocol.

The work that we presented focussed on Born-Markov master equations in Lindblad form. Preliminary studies indicate that the use of an extended Hilbert space, where the system strongly couples to an auxiliary, dissipative system \cite{Tamascelli:2018,Menu_etal22,Brown_etal-Kamal:2022}, could significantly increase the efficiency of the transfer. Future works will aim to perform an optimization in this regime, and assess the ultimate limit to the timescale of adiabatic transfer in an open-system setting.


\section*{Acknowledgments}
The authors are grateful to Christiane Koch and Josias Langbehn for discussions and helpful comments. This work was funded by the Deutsche Forschungsgemeinschaft (DFG, German Research Foundation) -- Project-ID 429529648 -- TRR306 QuCoLiMa (``Quantum Cooperativity of Light and Matter''), and by the German Ministry of Education and Research (BMBF) via the Project NiQ (Noise in Quantum Algorithms). L.\ G.\ acknowledges the QuantERA grant SiUCs (Grant No.\ 731473) and the PNRR MUR project PE0000023-NQSTI. This research was supported in part by the National Science Foundation under Grants No. NSF PHY-1748958 and PHY-2309135.

\section*{Numerical methods: implementation and availability of source code}
To provide a thorough analysis of the speed limits of the STARE protocol, see Sec.~\ref{s:speed_limit}, we extended our study to go beyond the Born-Markov limit. This investigation utilized a numerical simulation of a physical model of a qubit coupled to a damped auxiliary qubit. The source code and documentation to reproduce the results are available online at the link \url{https://zenodo.org/records/11058607}, see Ref.~\cite{king_2024_11058607}.

\bibliographystyle{quantum}
\bibliography{EK}


\onecolumn\newpage
\appendix 

\section{{Optimal transfer time} for the linear sweep function}
\label{app:LZ}
We analyze the scaling behavior of the transfer time when a linear sweep is implemented in the two-level system Hamiltonian \eqref{eq:LZ_Hamiltonian}. The analysis accounts for both purely unitary dynamics and incoherent dynamics with dephasing rate $\gamma$. Following Sec.~\ref{ss:LZ_and_RC_sweeps}, the sweep function is $s(t) = \epsilon t$, $\epsilon=(s_f-s_i)/T$, leading to a linear evolution schedule, $q_\mathrm{LZ}(\tau)=\tau$, in dimensionless coordinates. In contrast with the optimized schedules, $a$ \eqref{eq:dimensionless_parameters_a} and $b$ \eqref{eq:dimensionless_parameters_b} are no longer the natural parameters. Instead, we introduce
\begin{equation}
    \mathcal{A}^2 = \frac{a}{d_f-d_i} =\frac{g_0^2}{\epsilon} \quad \text{and} \quad  \mathcal{B} = \frac{b}{a} = \frac{\gamma}{g_0}\,.
\end{equation}
While keeping the dimensionless parameters $\mathcal{A}$ and $\mathcal{B}$ fixed, we take the limit $d_f=-d_i\rightarrow\infty$, yielding the infidelity
\begin{equation}
    \mathcal{I}(\mathcal{A},\mathcal{B})=\lim_{d_f\rightarrow\infty} \mathcal{I}(\mathcal{A},\mathcal{B}, -d_f,d_f)\,.
\end{equation}
Note that keeping $\mathcal{A}$ fixed requires that the transfer time $T$ scales as $1/g_0^2$. This is the standard Landau-Zener result, see Eq.~\eqref{eq:LZ_formula}. Consequently, if the minimum energy gap scales as $g_0\sim1/\sqrt{N}$, then $T\sim N$ and the performance in terms of transfer time is no better than the classical case. 

Let us now consider the STARE Lindbladian with the linear sweep. For $\mathcal{B}\neq 0$ and $\mathcal{B}$ to be constant, we should scale $g_0$ and $\gamma$ simultaneously in an identical manner. This behavior is neatly captured in the following expression:
\begin{equation}
    \mathcal{I}(\kappa^2 T, g_0/\kappa,\gamma/\kappa) = \mathcal{I}(T,g_0,\gamma)\,,\quad \kappa>0\,, 
\end{equation}
and confirmed, up to order $\mathcal{O}(T^{-2})$, by the analytic result for $\mathcal{I}(\mathcal{A},\mathcal{B})$ that was derived in Ref.~\cite{Avron_etal11_LZ}. Now to substantiate that the linear schedule does not lead to an optimal scaling of the transfer time, we consider two asymptotic regimes: moderate dephasing where $\gamma\ll g_0$ and dominant dephasing with $g_0\ll \gamma$. In the former case, we find \cite{Avron_etal11_LZ}
\begin{equation}\label{eq:Ilinear_analytic_weak_dephasing}
    \mathcal{I}\approx \frac{3\pi}{16}\mathcal{A}^{-2} \mathcal{B}\,,
\end{equation}
while for strong dephasing the infidelity is approximated by \cite{Avron_etal11_LZ}
\begin{equation}\label{eq:Ilinear_analytic_strong_dephasing}
    \mathcal{I}\approx \frac{\pi}{4}\mathcal{A}^{-2}\mathcal{B}^{-1}\,.
\end{equation}
If $\gamma\sim g_0$, as in the case of the optimized schedules, we obtain $T\sim N$ and the quadratic speed-up is lost. This serves to highlight the sensitivity of the transfer time scaling behavior to the choice of schedule $q$. 

\section{Power series expansion coefficients of the density matrix} \label{app:density_matrix_expansion}
In this appendix, we systematically compute each coefficient of Eq.~\eqref{eq:recursive_relation} to produce the result in Eq.~\eqref{eq:list_of_density_matrix_coeff_a_and_b} of the main text. Note that enforcing the initial condition $\varrho(0)=P_-(0)$ requires $a_0(0)=P_-(0)$ and $a_1(0)=a_2(0)=0$. This leads to the following coefficients, which are required to produce an expansion to $\mathcal{O}(\lambda_0^{-2})$, with $\lambda_0 \equiv T\sqrt{g_0^2 +\gamma^2}$:
\begin{align}
        a_0(\tau) &=  W(\tau,0)P_-(0) = P_-(\tau)\,,\\
\label{eq:rho_expansion_coefficient_b_1}
        b_1(\tau) &= \mathcal{L}^{-1}(\tau) \dot{\mathcal{P}}(\tau) a_0(\tau)\,,\\
\label{eq:rho_expansion_coefficient_a_1}
        a_1(\tau) &= \int_0^{\tau}d\tau'\, W(\tau,\tau')\dot{\mathcal{P}}(\tau')b_1(\tau')\,,\\
\label{eq:rho_expansion_coefficient_b_2}
        b_2(\tau) &= \mathcal{L}^{-1}(\tau) \dot{\mathcal{P}}(\tau) a_1(\tau) + \mathcal{L}^{-1}(\tau) \mathcal{Q}(\tau) \dot{b}_1(\tau)\,,\\
\label{eq:rho_expansion_coefficient_a_2}
        a_2(\tau) &= \int_0^{\tau}d\tau'\, W(\tau,\tau')\dot{\mathcal{P}}(\tau')b_2(\tau')\,.
\end{align}

The first coefficient, $a_0(\tau) = P_-(\tau)$, is trivially determined using the intertwining property of the propagator, see Ref.~\cite{Fraas_Hanggli17}. To calculate $b_1$, we insert the result for $a_0$ and note that $\mathcal{P}\dot{P}_\pm=0$, as well as $P_\pm \dot{P}_\pm P_\pm=0$. This leads to
\begin{equation}\label{eq:PdotActingOnPminus}
    \dot{\mathcal{P}}(\tau) P_-(\tau) = \frac{d}{d\tau}\left(\mathcal{P}(\tau)P_-(\tau)\right) = \dot{P}_-(\tau)\,,
\end{equation}
where we have used definition \eqref{eq:PandQ_superprojectors} for the superprojector $\mathcal{P}$ acting on the spectral projection $P_-$ of the Hamiltonian. Before the inverse Lindbladian acts on the result in Eq.~\eqref{eq:PdotActingOnPminus}, we manipulate the expression to obtain
\begin{align}
    \dot{P}_- = 2\dot{P}_- -\dot{P}_- &= 2\dot{P}_- - P_-\dot{P}_- - \dot{P}_-P_-\nonumber\\
    &=(\mathbb{I}-P_-)\dot{P}_- + \dot{P}_-(\mathbb{I}-P_-)\nonumber\\
    &=P_+ \dot{P}_-  + \dot{P}_-  P_+\,.\label{eq:Pdot_minus_identity}
\end{align}
Combining the result above with $\dot{P}_- =\frac{d}{d\tau}(|-\rangle\langle-|)=|\dot{-}\rangle\langle-| + |-\rangle\langle\dot{-}|$, we observe that $\dot{P}_-(\tau)$ in Eq.~\eqref{eq:PdotActingOnPminus} can be expressed as
\begin{equation}
    \dot{P}_-(\tau) = \langle+|\dot{-}\rangle\,|+\rangle\langle-| + \langle\dot{-}|+\rangle\,|-\rangle\langle+|
    = \langle+|\dot{-}\rangle\, S_{+,-} + \langle\dot{-}|+\rangle\, S_{-,+} \,,
\end{equation}
where $S_{ij}=|i\rangle\langle j|$, $i,j\in\{+,-\}$, are the right eigenvectors of $\mathcal{L}$. Since the Lindbladian's inverse in Eq.\ \eqref{eq:rho_expansion_coefficient_b_1} is defined on $\text{ran}(\mathcal{L})$, the kernel only contains the null vector, hence $\mathcal{L}$ is invertible. The eigenvalue equation $\mathcal{L}S_{ij}=\lambda_{ij}S_{ij}$, with $i\neq j$, can therefore be rewritten as $\mathcal{L}^{-1}S_{ij}=\lambda_{ij}^{-1} S_{ij}$. In addition, we can infer that $\dot{P}_-\in \text{ran}(\mathcal{L})$ from the identity $\mathcal{P}\dot{P}_\pm=0$, and it follows that 
\begin{equation}\label{eq:Inverse_L_on_PdotMin}
    \mathcal{L}^{-1}(\tau) \dot{P}_-(\tau) = \frac{1}{\lambda_{+,-}} \langle+|\dot{-}\rangle\, S_{+,-} + \frac{1}{\lambda_{-,+}}\langle\dot{-}|+  \rangle\, S_{-,+}\,. 
\end{equation}
Using the shorthand notation $\lambda_+\equiv \lambda_{+,-}$ and $\lambda_-\equiv \lambda_{-,+}$ of the main text, the resulting expansion coefficient is
\begin{equation}\label{eq:rho_expansion_coefficient_b_1_FINAL}
    b_1(\tau) = \frac{P_+(\tau) \dot{P}_- (\tau)}{\lambda_{+}(\tau)} + \frac{\dot{P}_-(\tau)  P_+(\tau)}{\lambda_{-}(\tau)}\,,
\end{equation}
 with $\lambda_{\pm}$ the time-dependent (dimensionless) eigenvalues of the STARE Lindbladian, see Eq.~\eqref{eq:Lindbladian_eigenvalues_MAIN}.
 
 We proceed by calculating the coefficient $a_1(\tau)$, which is given in Eq.~\eqref{eq:rho_expansion_coefficient_a_1}. Substituting in the result for $b_1(\tau)$~\eqref{eq:rho_expansion_coefficient_b_1_FINAL}, we obtain 
 \begin{equation}\label{eq:a_1_intermediate}
    a_1(\tau) = \int_0^{\tau}d\tau'\, W(\tau,\tau')\dot{\mathcal{P}}(\tau')\left[\frac{P_+(\tau') \dot{P}_- (\tau')}{\lambda_{+}(\tau')} + \frac{\dot{P}_-(\tau')  P_+(\tau')}{\lambda_{-}(\tau')}\right].
\end{equation}
Using the projector identities 
\begin{equation} \label{eq:projector_identity}
    P_\pm \ddot{P}_\mp P_\pm = 2 P_\pm \dot{P}_\mp^2P_\pm \qquad \text{and}\qquad P_\pm \ddot{P}_\pm P_\pm = -2 P_\pm \dot{P}_\pm^2P_\pm\,,
\end{equation}
as well as the fact that $\mathcal{P}P_+\dot{P}_-=0$, the superprojecter $\dot{\mathcal{P}}$ \eqref{eq:PandQ_superprojectors} acts on the combination of projectors $P_+\dot{P}_-$ to give
\begin{equation}\label{eq:projector_onto_Pplus_Pmindot}
    \dot{\mathcal{P}} (P_+\dot{P}_-) = \frac{d}{d\tau}(\mathcal{P}P_+\dot{P}_-)-\mathcal{P}\dot{P}_+\dot{P}_--\mathcal{P}P_+\ddot{P}_-
    = \mathcal{P}\dot{P}_-^2 - \mathcal{P}P_+\ddot{P}_-
    = P_- \dot{P}_-^2 P_- - P_+\dot{P}_+^2 P_+\,.
\end{equation}
Now using Eq.~\eqref{eq:Pdot_minus_identity} we recast the projectors in the second term of Eq.~\eqref{eq:a_1_intermediate} as $\dot{P}_- P_+ = \dot{P}_- - P_+ \dot{P}_-$. The result for $\dot{\mathcal{P}}(P_+\dot{P}_-)$ has already been calculated in Eq.~\eqref{eq:projector_onto_Pplus_Pmindot}, therefore we continue by calculating $\dot{\mathcal{P}}\dot{P}_-$:
\begin{equation}\label{eq:projector_onto_Pmindot}
    \dot{\mathcal{P}}\dot{P}_- = \frac{d}{d\tau} (\mathcal{P}\dot{P}_- ) - \mathcal{P}\ddot{P}_- 
    = -P_- \ddot{P}_- P_- - -P_+ \ddot{P}_- P_+
    = 2(P_- \dot{P}_-^2 P_- - P_+\dot{P}_+^2 P_+)\,,
\end{equation}
where we exploited the condition $\mathcal{P}\dot{P}_-=0$ and applied identity \eqref{eq:projector_identity} in the second step. Implementing Eqs.~\eqref{eq:projector_onto_Pplus_Pmindot} and \eqref{eq:projector_onto_Pmindot}, the coefficient $a_1$ can be simplified to
\begin{equation}\label{eq:a_1_intermediate2}
    a_1(\tau) = \int_0^{\tau}d\tau'\,  \left(\frac{1}{\lambda_{+}(\tau')} +\frac{1}{\lambda_{-}(\tau')}\right)W(\tau,\tau')
      \left[P_-(\tau') \dot{P}_-^2(\tau') P_-(\tau') - P_+(\tau')\dot{P}_+^2(\tau') P_+(\tau')\right].
\end{equation}
The final step involves the application of the adiabatic evolution generator $W$ on the sequence of projectors. To simplify the calculation, we reformulate $W(\tau,\tau')$ \eqref{eq:propagator_parallel_transport} in the convenient form \cite{Albash_etal12}
\begin{equation} \label{eq:propagator_Sum_Definition}
    W(\tau,\tau') = \sum_{j=\pm} |j(\tau)\rangle\langle j(\tau')|\,,
\end{equation}
where $|\pm(\tau)\rangle$ are the eigenstates forming the adiabatic basis of the two-level system Hamiltonian \eqref{eq:LZ_Hamiltonian}. The propagator \eqref{eq:propagator_Sum_Definition} now acts on the projectors, leading to
\begin{subequations}\label{eq:W_acting_on_projectors}
    \begin{align}
    W(\tau,\tau')&\left[P_-(\tau') \dot{P}_-^2(\tau') P_-(\tau') - P_+(\tau')\dot{P}_+^2(\tau') P_+(\tau')\right] \\
    &= |-(\tau)\rangle\langle -(\tau')|\dot{P}_-^2(\tau') P_-(\tau') - |+(\tau)\rangle\langle +(\tau')|\dot{P}_+^2(\tau') P_+(\tau')\\
    &= \langle -(\tau')|\dot{P}_-^2(\tau')|-(\tau')\rangle\, |-(\tau)\rangle\langle -(\tau')| - \langle +(\tau')|\dot{P}_-^2(\tau')|+(\tau')\rangle\, |+(\tau)\rangle\langle +(\tau')|\\
    &= \text{Tr}\left\{P_+(\tau')\dot{P}_-^2(\tau')\right\} \left(|-(\tau)\rangle\langle -(\tau')| - |+(\tau)\rangle\langle +(\tau')|\right)\mathbb{I}\,, \quad \text{where} \quad \mathbb{I} = P_-(\tau') + P_+(\tau')\\
    &=\text{Tr}\left\{P_+(\tau')\dot{P}_-^2(\tau')\right\}\, \left[P_-(\tau) - P_+(\tau)\right].
\end{align}
\end{subequations}
Note that the trace enters via the expectation values, with $\langle -(\tau')|\dot{P}_-^2(\tau')|-(\tau')\rangle = \langle +(\tau')|\dot{P}_-^2(\tau')|+(\tau')\rangle = \text{Tr}\{P_+(\tau')\dot{P}_-^2(\tau')\}$. Inserting the final result of Eq.~\eqref{eq:W_acting_on_projectors} into Eq.~\eqref{eq:a_1_intermediate2} leads to 
 \begin{equation} \label{eq:rho_expansion_coefficient_a_1_FINAL}
     a_1(\tau) = [P_-(\tau)-P_+(\tau)] \mathcal{J}(\tau)\,,\quad \mathcal{J}(\tau)\equiv\int_0^\tau d\tau'\,  \frac{2\text{Re}[\lambda_{-}(\tau')]}{|\lambda_{-}(\tau')|^2} \,\text{Tr}\left\{P_+(\tau')[\dot{P}_-(\tau')]^2\right\} \leq 0\,.
 \end{equation}

The two remaining coefficients, $a_2$ and $b_2$, contribute to the term of quadratic order in $1/\lambda_0$ in the expansion of the density matrix, see Eq.~\eqref{eq:density_matrix_expansion_ito_a_b}. We start with the expansion coefficient $b_2(\tau)$ \eqref{eq:rho_expansion_coefficient_b_2}. The first term depends on the coefficient $a_1$ \eqref{eq:rho_expansion_coefficient_a_1_FINAL}, with its explicit form given by
\begin{equation}
    \mathcal{L}^{-1}(\tau) \dot{\mathcal{P}}(\tau) a_1(\tau) = \mathcal{J}(\tau) \mathcal{L}^{-1}(\tau) (\dot{P}_-(\tau) - \dot{P}_+(\tau)) = 2\mathcal{J} (\tau)\mathcal{L}^{-1}(\tau) \dot{P}_-(\tau) \,,
\end{equation}
where we have used result \eqref{eq:PdotActingOnPminus}. Using Eq.~\eqref{eq:Inverse_L_on_PdotMin}, we can simplify this directly to obtain
\begin{equation} \label{eq:b_2_first_term}
    \mathcal{L}^{-1}(\tau) \dot{\mathcal{P}}(\tau) a_1(\tau) = 2\mathcal{J}(\tau)  \left( \frac{P_+(\tau) \dot{P}_- (\tau)}{\lambda_{+}(\tau)} + \frac{\dot{P}_-(\tau)  P_+(\tau)}{\lambda_{-}(\tau)}\right).
\end{equation}
The second term of $b_2$ \eqref{eq:rho_expansion_coefficient_b_2} depends on the complement $\mathcal{Q}=\mathbb{I}-\mathcal{P}$ of superprojector $\mathcal{P}$, and the derivative of $b_1$ \eqref{eq:rho_expansion_coefficient_b_1_FINAL} with respect to $\tau$, where the latter is given by
\begin{equation}
    \dot{b}_1(\tau) = -\frac{\dot{\lambda}_{+}}{\lambda^2_{+}}P_+ \dot{P}_- -\frac{\dot{\lambda}_{-}}{\lambda^2_{-}}\dot{P}_-P_+ 
     + \frac{1}{\lambda_{+}} \left(P_+ \ddot{P}_- - \dot{P}_-^2\right)
    + \frac{1}{\lambda_{-}} \left( \ddot{P}_- P_+ - \dot{P}_-^2\right).
\end{equation}
Evaluating $[\dot{b}_1(\tau)-\mathcal{P}(\tau) \dot{b}_1(\tau)]$, and performing some manipulation of the projectors, results in
    \begin{subequations} \label{eq:b_1_minus_P_b_1}
        \begin{align}
            \dot{b}_1(\tau)-\mathcal{P}(\tau) \dot{b}_1(\tau) &= -\frac{\dot{\lambda}_{+}}{\lambda^2_{+}}P_+ \dot{P}_- -\frac{\dot{\lambda}_{-}}{\lambda^2_{-}}\dot{P}_-P_+ + \frac{1}{\lambda_{+}} \left( P_+ \ddot{P}_- - \dot{P}_-^2 - P_+\ddot{P}_-P_+ + P_+\dot{P}_-^2 P_+ + P_-\dot{P}_-^2 P_-\right) \nonumber\\
            &\qquad \quad + \frac{1}{\lambda_{-}} \left(  \ddot{P}_- P_+ - \dot{P}_-^2 - P_+\ddot{P}_-P_+ + P_+\dot{P}_-^2 P_+ + P_-\dot{P}_-^2 P_-\right)\\
            &=-\frac{\dot{\lambda}_{+}}{\lambda^2_{+}}P_+ \dot{P}_- -\frac{\dot{\lambda}_{-}}{\lambda^2_{-}}\dot{P}_-P_+ + \frac{1}{\lambda_{+}} P_+ \ddot{P}_- P_- + \frac{1}{\lambda_{-}} P_- \ddot{P}_- P_+ \,.
        \end{align}
    \end{subequations}
From previous calculations it is straightforward to show that the inverse Lindbladian $\mathcal{L}^{-1}$ acts on the first two terms of \eqref{eq:b_1_minus_P_b_1}, with $P_+ \dot{P}_- = \langle+|\dot{-}\rangle\, S_{+,-}$ and $\dot{P}_- P_+  = \langle\dot{-}|+\rangle\, S_{-,+}$, to give 
\begin{equation} \label{eq:b_1_minus_P_b_1_FIRST_TWO_TERMS}
    -\frac{\dot{\lambda}_{+}}{\lambda^3_{+}}P_+ \dot{P}_- -\frac{\dot{\lambda}_{-}}{\lambda^3_{-}}\dot{P}_-P_+\,.
\end{equation}
Noting that $\ddot{P}_-=\frac{d^2}{d\tau^2} (|-\rangle\langle-|)$ and $\frac{d}{d\tau}|\pm\rangle = |\dot{\pm}\rangle = \mp \langle +|\dot{-}\rangle |\mp\rangle$, it is now possible to show that 
\begin{equation}
    P_+ \ddot{P}_- P_- = \frac{d}{d\tau}(\langle +|\dot{-}\rangle) S_{+,-}\,, \quad P_- \ddot{P}_- P_+ = \frac{d}{d\tau}(\langle \dot{-}|+\rangle) S_{-,+}\,.
\end{equation}
Consequently, the inverse Lindbladian $\mathcal{L}^{-1}$ acts on the final two terms in Eq.~\eqref{eq:b_1_minus_P_b_1}, resulting in
\begin{equation} \label{eq:b_1_minus_P_b_1_SECOND_TWO_TERMS}
     +\frac{1}{\lambda^2_{+}} \frac{\frac{d}{d\tau}(\langle +|\dot{-}\rangle)}{\langle+|\dot{-}\rangle}P_+ \dot{P}_- +\frac{1}{\lambda^2_{-}} \frac{\frac{d}{d\tau}(\langle \dot{-}|+\rangle)}{\langle \dot{-}|+\rangle}\dot{P}_-P_+\,.
\end{equation}
Finally, defining
\begin{equation}\label{eq:x1}
    x_1(\tau) = \left(\frac{2\mathcal{J}(\tau)}{\lambda_{+}(\tau)}-\frac{\dot{\lambda}_{+}(\tau)}{\lambda^3_{+}(\tau)} +\frac{1}{\lambda^2_{+}(\tau)} \frac{\frac{d}{d\tau}(\langle +|\dot{-}\rangle)}{\langle+|\dot{-}\rangle(\tau)}\right)
\end{equation}
and
\begin{equation}\label{eq:x2}
    x_2(\tau) = \left(\frac{2\mathcal{J}(\tau)}{\lambda_{-}(\tau)}-\frac{\dot{\lambda}_{-}(\tau)}{\lambda^3_{-}(\tau)} +\frac{1}{\lambda^2_{-}(\tau)} \frac{\frac{d}{d\tau}(\langle \dot{-}|+\rangle)}{\langle \dot{-}|+\rangle(\tau)}\right),
\end{equation}
we combine the results in Eqs.~\eqref{eq:b_2_first_term}, \eqref{eq:b_1_minus_P_b_1_FIRST_TWO_TERMS} and \eqref{eq:b_1_minus_P_b_1_SECOND_TWO_TERMS} to acquire
\begin{equation} \label{eq:rho_expansion_coefficient_b_2_FINAL}
    b_2(\tau) =  x_1(\tau) P_+(\tau) \dot{P}_-(\tau) + x_2(\tau)\dot{P}_-(\tau) P_+(\tau)\,.
\end{equation}

To recover a simple expression for coefficient $a_2$ \eqref{eq:rho_expansion_coefficient_a_2}, we use Eq.~\eqref{eq:rho_expansion_coefficient_b_2_FINAL} and follow the same set of steps as for coefficient $a_1$. After some manipulation, $a_2$ takes the form
\begin{equation} \label{eq:rho_expansion_coefficient_a_2_FINAL}
      a_2(\tau) = [P_-(\tau)-P_+(\tau)] \int_0^{\tau}d\bar{\tau}\, \text{Tr}\left\{P_+(\bar{\tau})[\dot{P}_-(\bar{\tau})]^2\right\}    \big(x_1(\bar{\tau}) + x_2(\bar{\tau})\big)\,,
\end{equation}
with $x_{1}$ and $x_2$ defined in Eqs.~\eqref{eq:x1} and \eqref{eq:x2}, respectively. The sum of $x_{1}$ and $x_2$ can be simplified to
\begin{equation}
    x_1 + x_2 = \frac{\dot{q}d(d_f-d_i)}{T^2\left(\gamma ^2+\Delta^2\right)}\left[-\frac{8 \gamma^2 g_0^2 }{\left(\gamma ^2+\Delta^2\right)^2} + \frac{2  \left(g_0^2 \left(d^2+1\right)-4 \gamma^2\right)}{\left(d^2+1\right) \left(\gamma ^2+\Delta^2\right)} 
    + \frac{4}{\left(d^2+1\right) }
    - \frac{4 \gamma T \mathcal{J} }{\dot{q}d (d_f-d_i)  }\right],
\end{equation}
where $d=d(\tau)=s(\tau)/g_0$ is defined in Eq.~\eqref{eq:sweep_function_dimensionless_parameters}, with initial and final values $d_i$ and $d_f$, respectively, and $q$ is the evolution schedule. The dephasing strength $\gamma$ enters via the spectrum of the Lindbladian \eqref{eq:Lindbladian_eigenvalues_MAIN}, and the instantaneous energy gap $\Delta =\Delta(\tau) = E_+(\tau)-E_-(\tau)$ of the isolated two-level system comes from $\lambda_{\pm}$ \eqref{eq:Lindbladian_eigenvalues_MAIN} in Eqs.~\eqref{eq:x1} and \eqref{eq:x2}, with its minimum value denoted by $g_0$. Recall that $\mathcal{J}=\mathcal{J}(\tau)$ is the integral of  Eq.~\eqref{eq:rho_expansion_coefficient_a_1_FINAL}. As required, the integrand in Eq.~\eqref{eq:rho_expansion_coefficient_a_2_FINAL} is then dimensionless. Combining the results of all the coefficients, we arrive at the power series expansion \eqref{eq:density_matrix_expansion_ito_a_b} of the density matrix $\varrho$, with the expansion coefficients summarized in Eq.~\eqref{eq:list_of_density_matrix_coeff_a_and_b} of the main text.

\section{Derivation of the function $M(q)$ in the infidelity of the non-unitary protocol} \label{app:Avron_derivation}
In Sec.~\ref{s:adiabatic_Lindblad_dynamics} of the main text we provide an analytic expression for the infidelity $\mathcal{I}$ of adiabatic transfer for a non-unitary protocol with an evolution schedule $q$. To derive the closed-form expression of Eq.~\eqref{eq:infidelity_analytic_Avron}, we insert the expansion of $\varrho(1)$ from Eq.~\eqref{eq:density_matrix_expansion_ito_a_b} into $\mathcal{I} = 1-\text{Tr}\{P_-(1)\varrho(1)\}$. Note that due to the cyclic property of the trace and orthogonality of the basis states $\ket{\pm}_\tau$, implying that $P_+P_-=0$, the $b_1$ and $b_2$ terms in the expansion of Eq.~\eqref{eq:density_matrix_expansion_ito_a_b} vanish. The zeroth-order term $a_0$ trivially leads to 1 upon evaluating the trace, and the $a_2$ term, which is of order $1/\lambda_0^2$, is viewed as a correction term $\mathcal{C}$, see Eq.~\eqref{eq:C_q_function}. We therefore focus our attention on the contribution of the $\mathcal{O}(1/\lambda_0)$ term to the infidelity, with
\begin{equation}
    \mathcal{I} = -\text{Tr}\left\{P_-(1) [P_-(1)-P_+(1)]  \mathcal{J}(1) \right\} + \mathcal{O}(\lambda_0^{-2})\,.
\end{equation}
The function $\mathcal{J}$ may be extracted from the trace, leading to
\begin{equation}\label{eq:appInfitoA}
    \mathcal{I} = -\mathcal{J}(1) + \mathcal{O}(\lambda_0^{-2})\,.
\end{equation}
In Appendix \ref{app:density_matrix_expansion} we derived the integral expression for $\mathcal{J}(\tau)$ \eqref{eq:rho_expansion_coefficient_a_1_FINAL}, which is negative and has an integrand that takes the form $\mathcal{A}(\tau) = 2\text{Re}[\lambda_{-}(\tau)] \,\text{Tr}\left\{P_+(\tau)[\dot{P}_-(\tau)]^2\right\}/|\lambda_{-}(\tau)|^2$. Upon inserting the eigenvalues $\lambda_{\pm}(\tau)$ of Eq.~\eqref{eq:Lindbladian_eigenvalues_MAIN}, the integrand of $\mathcal{J}$ reads
\begin{equation}\label{eq:appAFunction}
    \mathcal{A}(\tau) = \frac{1}{T}\frac{-2\gamma(\tau) }{\gamma^2(\tau)+\Delta^2(\tau)} \,\text{Tr}\left\{P_+(\tau)[\dot{P}_-(\tau)]^2\right\}  \,.
\end{equation}
We perform the replacement $\dot{P}_- \rightarrow \dot{q} P_-'$, where the prime denotes the derivative with respect to the evolution schedule $q$, and evaluate the trace:
\begin{equation}
    \text{Tr}\left\{P_+(\tau)[P_-'(\tau)]^2 \right\} = \frac{g_0^2 (s(t_f)-s(t_i))^2}{4 \left(g_0^2+[q (s(t_f)-s(t_i))+s(t_i)]^2\right)^2}\,.
\end{equation}
Using the shorthand notation of Eq.~\eqref{eq:sweep_function_dimensionless_parameters}, the expression above reduces to the compact form \mbox{$(d_f-d_i)^2/[4(1+d^2(\tau))^2]$}. Combining this result with the prefactor of Eq.~\eqref{eq:appAFunction}, in which we recast the instantaneous energy gap $\Delta(\tau)$ as $[E_+(\tau)-E_-(\tau)]^2=g_0^2 \left(1+d^2(\tau)\right)$, we obtain
\begin{equation}\label{eq:appAFunction2}
    \mathcal{A}(\tau) = \frac{-2\gamma(\tau)\dot{q}^2 (d_f-d_i)^2}{4 T\left[\gamma^2(\tau)+g_0^2 \left(1+d^2(\tau)\right)\right] (1+d^2(\tau))^2}  \,.
\end{equation}
After inserting the function $\mathcal{J}(1) = \int_0^1 d\tau\,\mathcal{A}(\tau)$, with $\mathcal{A}$ of Eq.~\eqref{eq:appAFunction2}, into the expression for the infidelity \eqref{eq:appInfitoA}, we extract the factor $-2\dot{q}^2$, and write the minimum gap $g_0$ and dephasing rate $\gamma$ in terms of the dimensionless parameters $a$ \eqref{eq:dimensionless_parameters_a} and $b$ \eqref{eq:dimensionless_parameters_b}, respectively. This leads directly to the result stated in the main text:
\begin{equation} 
 	\mathcal{I} = 2 \int_0^1 d\tau\, M(q)\dot{q}^2 + \mathcal{O}(\lambda_0^{-2})\,, \quad M(q) = \frac{b  (d_f-d_i)^2}{4 \left(1+d^2\right)^2 \left(a^2 \left(1+d^2\right)+b^2\right)}\geq 0\,,
\end{equation}
where the $q$-dependence of the function $M$ enters via the parameter $d$. It can be shown that this result is a special case of the more general result reported in Ref.~\cite{Avron_etal10}.

\section{Adiabatic Markovian master equation microscopic derivation}\label{app:microscopic_derivation}
In this appendix, we outline the microscopic derivation of the adiabatic, Markovian, STARE master equation with the Lindbladian of Eq.~\eqref{eq:dephasingME_two_level}. To arrive at the STARE master equation from a physical microscopic model, we require several approximations to be made and must impose constraints on various parameters to satisfy the adiabatic condition. In Sec.~\ref{app:BM_approx} we discuss these approximations, which are typically grouped together and collectively referred to as the Born-Markov approximation \cite{Breuer_Petruccione07}. This is followed by the application of the adiabatic theorem in Sec.~\ref{app:AL}, which closely follows Ref.~\cite{Albash_etal12}. We close this appendix with Sec.~\ref{sss:full_ME_result}, which combines the results of the previous sections to arrive at the adiabatic Markovian Lindblad master equation utilized in the main text.

\subsection{Born-Markov approximation} \label{app:BM_approx}
 In this section, we will apply the Born-Markov approximation to reduce the von Neumann equation to an integro-differential equation for the reduced density matrix $\varrho$. To separate the interaction dynamics from the free dynamics, we continue by working in the interaction picture, where operators and density matrices are distinguished from their Schr\"{o}dinger picture counterparts by means of a tilde $\Tilde{\blacksquare}$. To transform to the interaction picture, we define the unitary evolution operator $U_0(t,t')$ of the free system and reservoir as
$U_0(t,t') = U_\text{S}(t,t')\otimes U_\text{R}(t,t')$, where 
\begin{equation}\label{eq:unitary_evolution_operator_free_system}
    U_\text{S}(t,t') = \mathcal{T}_\leftarrow \exp\left(-i\int_{t'}^td\tau\, H_\text{S}(\tau) \right)
\end{equation}
acts on the system and depends on the time-ordering operator $\mathcal{T}_\leftarrow$, while 
\begin{equation}\label{eq:unitary_evolution_operator_reservoir}
    U_\text{R}(t,t') = \exp\left(-iH_\text{R}(t-t') \right)
\end{equation}
acts on the reservoir. Similarly, the composite system-reservoir evolution operator $U_\text{SR}(t,t')$ is obtained by solving the time-dependent Schr\"{o}dinger equation, but now in the interacting case, giving
\begin{equation} \label{eq:unitary_evolution_operator_composite_system}
    U_\text{SR}(t,t') = \mathcal{T}_\leftarrow \exp\left(-i\int_{t'}^td\tau\, H_\text{SR}(\tau) \right).
\end{equation}
It follows that $\tilde{U}_\text{SR}(t,0) = U_0^\dag(t,0) U_\text{SR}(t,0)$. For all other operators, as well as the density matrix, we relate the original Schr\"{o}dinger picture variable $\blacksquare$ to the interaction picture via
\begin{equation}\label{eq:SP_to_IP}
    \tilde{\blacksquare}(t) = U_0^\dag(t,0) \blacksquare U_0(t,0)\,.
\end{equation}
This enables us to write the von Neumann equation in terms of the incoherent contribution to the dynamics as \cite{Breuer_Petruccione07,Rivas_Huelga}
\begin{equation} \label{eq:von_Neumann_eq_interaction_picture}
    \frac{d}{dt}\tilde{\varrho}_\text{SR}(t) = -i[\tilde{H}_\text{I}(t),\tilde{\varrho}_\text{SR}(t)]\,, \quad \tilde{\varrho}_\text{SR}(0) = \varrho_\text{SR}(0)\,,
\end{equation}
with the solution \cite{Breuer_Petruccione07,Albash_etal12}
\begin{equation}\label{eq:sol_to_von_Neumann_in_IP}
    \tilde{\varrho}_\text{SR}(t) = \tilde{\varrho}_\text{SR}(0) - i\int_0^t d\tau\,\left[ \tilde{H}_\text{I}(\tau),\tilde{\varrho}_\text{SR}(\tau)\right].
\end{equation}
Note that Eq.~\eqref{eq:von_Neumann_eq_interaction_picture} can be expanded self-consistently to an arbitrary order by inserting solution \eqref{eq:sol_to_von_Neumann_in_IP}.

Now, we outline three assumptions that will be imposed in the remaining steps \cite{Breuer_Petruccione07,Rivas_Huelga}:
\begin{enumerate}[i.]
    \item \textit{Born approximation.} The initial correlations between the system and reservoir are negligible such that $\tilde{\varrho}_\text{SR}=\tilde{\varrho}(t)\otimes \tilde{\varrho}_\text{R}(t) + \chi(t)$, where $\chi(t)\approx 0$.
    \item \textit{Weak-coupling limit.} The coupling $x_0$ is adequately weak to leave the state of the reservoir globally constant over time, i.e.\ $\tilde{\varrho}_\text{SR}\approx \tilde{\varrho}(t)\otimes \varrho_\text{R}$.
    \item \textit{Markov approximation.} There exists a sufficiently large timescale separation between the times over which the system and reservoir evolve, denoted by $\tau_\mathrm{S}$ and $\tau_\mathrm{R}$, respectively, with the dynamics of the former being much slower than the dynamics of the latter, i.e.\ $\tau_\mathrm{S}\gg \tau_\mathrm{R}$.
\end{enumerate}
The consequences of these assumptions are discussed in Sec.~\ref{s:speed_limit} of the main text, together with the restrictions they impose on certain parameters and timescales, such as $x_0$, $\tau_\mathrm{S}$ and $\tau_\mathrm{R}$. Now we expand the equation of motion \eqref{eq:von_Neumann_eq_interaction_picture} to first order, perform a partial trace over the reservoir degrees of freedom and neglect the inhomogeneous contribution, leading to \cite{Breuer_Petruccione07,Albash_etal12,Rivas_Huelga}
\begin{equation}
    \frac{d}{dt}\tilde{\varrho}(t) = -\int_0^t d\tau\, \text{Tr}_\text{R}\left\{\left[\tilde{H}_\text{I}(t), [\tilde{H}_\text{I}(t-\tau),\tilde{\varrho}_\text{SR}(t-\tau)]\right]\right\}.
\end{equation}
The shifted time has been introduced through a change of variables. After inserting the interaction Hamiltonian $\tilde{H}_\text{I}$, see the final term of Eq.~\eqref{eq:interaction_Hamiltonian_for_ME_general} which is recast in the interaction picture via Eq.~\eqref{eq:SP_to_IP}, and carrying out assumption (i) and (ii) above, we find \cite{Breuer_Petruccione07,Albash_etal12,Rivas_Huelga}
\begin{equation}\label{eq:ME_after_Born_Approx}
     \frac{d}{dt}\tilde{\varrho}(t) = \dot{\tilde{\varrho}}(t) = -\int_0^t d\tau\, \left\{\tilde{H}_\text{S}(t) \tilde{H}_\text{S}(t-\tau) \tilde{\varrho}(t-\tau)  -  \tilde{H}_\text{S}(t-\tau)\tilde{\varrho}(t-\tau) \tilde{H}_\text{S}(t)  \right\} C_\text{R}(t,t-\tau) + \text{h.c.}\,,
\end{equation}
with the reservoir autocorrelation functions
\begin{equation}\label{eq:correlation_function_general}
    C_\text{R}(t,t-\tau) \equiv x_0^2\,\text{Tr}_\text{R}\left\{\tilde{X}_\text{R}(t)\tilde{X}_\text{R}(t-\tau)\varrho_\text{R} \right\} 
    = x_0^2\left\langle \tilde{X}_\text{R}(t)\tilde{X}_\text{R}(t-\tau)\right\rangle.
\end{equation}
The stationarity of the reservoir, see assumption (ii), imposes a time homogeneity in the correlation functions such that $C_\text{R}(\tau,0)=C_\text{R}^{*}(0,\tau)$. Therefore, we denote the time homogeneous correlation functions by $C_\text{R}(\tau)$ for notational simplicity. Assuming Markovian dynamics, we consider the density matrix to be roughly constant over the interval $\tau\in[0,t]$, permitting the replacement $\tilde{\varrho}(t-\tau) \rightarrow \tilde{\varrho}(t)$ in Eq.~\eqref{eq:ME_after_Born_Approx}, as well as the extension of the integral's upper bound \cite{Breuer_Petruccione07,Albash_etal12,Rivas_Huelga}:
\begin{equation}\label{eq:integro_diff_eq}
      \dot{\tilde{\varrho}}(t) = \int_0^\infty d\tau\, \left\{\tilde{H}_\text{S}(t-\tau)\tilde{\varrho}(t-\tau) \tilde{H}_\text{S}(t)  
     -  \tilde{H}_\text{S}(t) \tilde{H}_\text{S}(t-\tau) \tilde{\varrho}(t-\tau) \right\} C_\text{R}(\tau) + \text{h.c.}
\end{equation}
This is in line with assumption (iii).  

\subsection{Adiabatic limit} \label{app:AL}
Our goal is to further simplify the integro-differential equation \eqref{eq:integro_diff_eq} by requiring that the dynamics satisfy the adiabatic condition. Introducing this adiabatic ``limit''  ensures that the coherent part of the dynamics remains adiabatic, while warranting the approximation of the system time evolution operator of Eq.~\eqref{eq:unitary_evolution_operator_free_system} by its adiabatic counterpart $U_\text{S}^\text{ad}(t,t')$. The approximation $U_\text{S}(t,t')\approx U_\text{S}^\text{ad}(t,t')$ provides a good description of the evolution up to $\mathcal{O}(h/g_0^2)$; see Ref.~\cite{Albash_etal12} for an extensive discussion. Below, we summarize the derivation of Ref.~\cite{Albash_etal12} for $U_\text{S}^\text{ad}(t,t')$ and then apply it to Eq.~\eqref{eq:integro_diff_eq}.

The rigorous treatment in Ref.~\cite{Albash_etal12} starts with the adiabatic propagator
\begin{equation}
    W(t,t') = \sum_{j=\pm} |j(t)\rangle\langle j(t')|\,,
\end{equation}
which was inspired by Kato's formulation of the adiabatic theorem \cite{Kato50} and analyzed extensively by Avron \textit{et al}.~\cite{Avron_etal87}. Roughly speaking, $W$ evolves an instantaneous eigenstate $|j(t')\rangle$ of the system at time $t'$ to the corresponding eigenstate $|j(t)\rangle$ at a later time $t$. Moving to the so-called \textit{adiabatic interaction picture}, we apply the propagator $W$ to $U_\text{S}(t,t')$ \eqref{eq:unitary_evolution_operator_free_system},
\begin{equation}
    \bar{V}(t,t') = W^\dag(t,t') U_\text{S}(t,t')\,,
\end{equation}
where the overscore, which signifies that an operator is in the \textit{adiabatic} interaction picture, should not be confused with the tilde notation reserved for the standard interaction picture. Then it can be shown that $\bar{V}$ obeys the Schr\"{o}dinger equation $d\bar{V}(t,t')/dt = -i \bar{H}_\text{S}^\text{ad}(t,t') \bar{V}(t,t')$, where $\bar{H}_\text{S}^\text{ad} = \bar{H}_\text{S} - i W^\dag(t,t')[\dot{P}_-(t),P_-(t)]W(t,t')$ \cite{Albash_etal12,Avron_etal87}. Since the commutator in $\bar{H}_\text{S}^\text{ad}$ is the generator for purely adiabatic evolution \cite{Avron_etal87}, $\bar{H}_\text{S}^\text{ad}$ acts as a perturbation in the adiabatic limit. Hence, we solve the Schr\"{o}dinger equation using perturbation theory, leading to \cite{Albash_etal12}
\begin{equation}\label{eq:V_perturbative_expansion}
    \bar{V}(t,t') = \bar{V}_0(t,t') \left[\mathbb{I} + \bar{V}_1(t,t')+\ldots\right],\quad \bar{V}_0(t,t')= \mathcal{T}_\leftarrow \exp\left(-i\int_{t'}^t d\tau\,(\bar{H}_\text{S}(t,t')-\bar{H}_\text{G}(t,t'))\right).
\end{equation}
The system Hamiltonian is given in Eq.~\eqref{eq:LZ_Hamiltonian}, while $H_\text{G}(t)=\sum_a\phi_a(t) P_a(t)$, with Berry connection $\phi_a(t)=i\leftindex_t{}\langle a|\dot{a}\rangle_t$, accounts for the geometric phase \cite{Albash_etal12}. We only consider the leading-order term $\bar{V}_0(t,t')$ in perturbation theory, giving rise to the adiabatic evolution operator
\begin{equation}\label{eq:adiabatic_evolution_operator_free_system}
    U_\text{S}^\text{ad}(t,t') = W(t,t')\Bar{V}_0(t,t') = \sum_{j=\pm} |j(t)\rangle\langle j(t')| e^{-i\mu_j(t,t')},
\end{equation}
where the phase is $\mu_j(t,t')=\int_{t'}^td\tau\,[E_j(\tau)-\phi_j(\tau)]$.

\subsection{Full adiabatic Markovian master equation}\label{sss:full_ME_result}
To arrive at the full adiabatic, Markovian, STARE master equation, we combine the results of the previous two sections. First, recall that the unitary evolution operators $U_\text{S}(t-\tau,0)=U^\dag_\text{S}(t,t-\tau)U_\text{S}(t,0)$ and $U_\text{S}(t,0)$ enter via the interaction picture Hamiltonians $\tilde{H}_\text{S}$ in Eq.~\eqref{eq:integro_diff_eq}. In the adiabatic limit, we can perform the replacement $U_\text{S}(t,0)\rightarrow U_\text{S}^\text{ad}(t,0)$. Furthermore, consistency with the Markov approximation necessitates short-lived reservoir correlations, inferring that the system Hamiltonian is approximately time-independent over the interval $[t-\tau,t]$. We therefore take $U_\text{S}(t-\tau,0) \approx e^{iH_\text{S}(t)\tau} U_\text{S}^\text{ad}(t,0)$ in the integro-differential equation of Eq.~\eqref{eq:integro_diff_eq}. After some manipulation, we arrive at 
\begin{equation}\label{eq:rho_dot_IP}
    \dot{\tilde{\varrho}}(t) =\sum_{\mathclap{a,b=\pm}}\Gamma(0) E_a(t) E_b(t) P_a(0) \left[ \tilde{\varrho}(t),P_b(0)\right] +\text{h.c.}\,,
\end{equation}
where projections onto the instantaneous eigenstates of the two-level system are denoted by $P_\pm$, and
\begin{equation} \label{eq:spectral_density_ito_C_R}
   \Gamma(\omega) = \int_0^\infty d\tau\, e^{i\omega\tau} C_\text{R}(\tau)
\end{equation}
is the reservoir spectral density. For convenience, we replace the one-sided Fourier transform by a complete Fourier transform \cite{Albash_etal12}
\begin{equation} \label{eq:spectral_density_full_FT}
    \Gamma(\omega) = \Gamma_\text{re}(\omega)/2 + i \Gamma_\text{im}(\omega)\,,
\end{equation}
where the real and imaginary components are\footnote{In this context, $\mathcal{P}$ refers to the Cauchy principal value.}
\begin{equation}\label{eq:spectral_density_Re}
     \Gamma_\text{re}(\omega) = \int_{-\infty}^\infty d\tau\, e^{i\omega\tau} C_\text{R}(\tau) \qquad \text{and} \qquad \Gamma_\text{im}(\omega) = \frac{1}{2\pi}\int_{-\infty}^\infty d\omega'\, \Gamma_\text{re}(\omega') \mathcal{P}\left(\frac{1}{\omega-\omega'}\right).
\end{equation}
Finally, we rewrite Eq.~\eqref{eq:rho_dot_IP} in the Schr\"{o}dinger picture using relation \eqref{eq:SP_to_IP}, resulting in a quantum master equation in Lindblad form:
\begin{equation} \label{eq:final_ME_from_microscopic_derivation}
    \dot{\varrho}(t)=\mathcal{L}\varrho(t)=-i[H_\text{S}(t) + H_\text{LS}(t),\varrho(t)] + \mathcal{D}_\text{ad}(\varrho(t))
\end{equation}
with the adiabatic dissipative superoperator
\begin{equation} \label{eq:final_dissipator_from_microscopic_derivation}
    \mathcal{D}_\text{ad}(\varrho(t))= \Gamma_\text{re}(0)\sum_{\mathclap{a,b=\pm}} E_a(t)E_b(t) \left[P_a(t)\varrho(t)P_b(t) - \frac{1}{2}\{P_a(t)P_b(t),\varrho(t)\}\right].
\end{equation}
The Lamb-shift Hamiltonian $H_\text{LS}$ in master equation \eqref{eq:final_ME_from_microscopic_derivation} is
\begin{equation}
    H_\text{LS}(t)= \Gamma_\text{im}(0) \sum_{\mathclap{a,b=\pm}} E_a(t) E_b(t)  \left( [P_b(t),\varrho(t)]P_a(t) -P_a(t)[\varrho(t),P_b(t)]\right).
\end{equation}
Independent of the choice of reservoir, i.e.\ for any spectral density $\Gamma_\text{im}(0)$, the Lamb-shift Hamiltonian will vanish. This can be shown using the completeness relation for the projectors. 

\end{document}